\documentclass[conference]{IEEEtran}
\IEEEoverridecommandlockouts
\usepackage{cite}
\usepackage{amsmath,amssymb,amsfonts}
\usepackage{algorithmic}
\usepackage{graphicx}
\usepackage{textcomp}
\usepackage{xcolor}
\usepackage{url}
\usepackage{latexsym}
\usepackage{subfigmat}
\usepackage{comment}
\usepackage{multicol}
\usepackage{multirow}
\usepackage{figsize}

\def\BibTeX{{\rm B\kern-.05em{\sc i\kern-.025em b}\kern-.08em
    T\kern-.1667em\lower.7ex\hbox{E}\kern-.125emX}}

\begin{document}

\title{
%Backdoor-Assisted Membership Inference Attack
Do Backdoors Assist Membership Inference Attacks?
}

\author{\IEEEauthorblockN{1\textsuperscript{st} Yumeki Goto}
\IEEEauthorblockA{\textit{Graduate School of Information Science and Technology} \\
\textit{Osaka University}\\ Suita, Japan \\email address or ORCID}
\and
\IEEEauthorblockN{2\textsuperscript{nd} Given Name Surname}
\IEEEauthorblockA{\textit{dept. name of organization (of Aff.)} \\
\textit{name of organization (of Aff.)}\\ City, Country \\ email address or ORCID}
\and
\IEEEauthorblockN{3\textsuperscript{rd} Given Name Surname}
\IEEEauthorblockA{\textit{dept. name of organization (of Aff.)} \\
\textit{name of organization (of Aff.)}\\ City, Country \\ email address or ORCID}
}
\author{
\IEEEauthorblockN{
Yumeki Goto\IEEEauthorrefmark{1}, 
Nami Ashizawa\IEEEauthorrefmark{2}, 
Toshiki Shibahara\IEEEauthorrefmark{2},
Naoto Yanai\IEEEauthorrefmark{1} 
}
% Nami Ashizawa
% Toshiki Shibahara
% Naoto Yanai
% Yumeki Goto
\IEEEauthorblockA{
\IEEEauthorrefmark{1}
 Osaka University, 1-5 Ymadagaoka, Suita-shi, Osaka, 565-0871, Japan.
}
\IEEEauthorblockA{
\IEEEauthorrefmark{2}
 NTT Social Informatics Laboratories\\ 3-9-11 Midori-cho, Musashino-shi, Tokyo, 180-8585, Japan.
}
}
% \author{
% Yumeki Goto
%   後藤 勇芽輝
% \thanks {
%  Osaka University, 
% }
%   \thanks{
%     大阪大学, 〒565-0871 大阪府吹田市山田丘1-5 \\
%     Osaka University, 1-5 Ymadagaoka, Suita-shi, Osaka, 565-0871, Japan.y-goto@ist.osaka-u.ac.jp
%   }\\
%   Yumeki Goto
%   \and
%   芦澤 奈実             %和文の第一著者名
%   \thanks{
%     NTT社会情報研究所, 180-8585, 東京都武蔵野市緑町3-9-11\\         %和文の所属と住所
%     NTT Social Informatics Laboratories\\ 3-9-11 Midori-cho, Musashino-shi, 
%     Tokyo 180-8585    %英文の所属と住所
%     %(ご希望の場合は, 続けて電子メールアドレスを記載)
%     }\\
%   Nami Ashizawa          %英文の第一著者名
%   \and
%   矢内 直人             %和文の第二著者名
%   \samethanks{1}\\
%   Naoto Yanai         %英文で第二著者名
%   \and
%   芝原 俊樹             %和文の第二著者名
%   \samethanks{2}\\
%   Toshiki Shibahara         %英文で第二著者名
%   \and
% }

\maketitle

%page limitation: 6 pages

\begin{abstract}
When an adversary provides poison samples to a machine learning model, privacy leakage, such as membership inference attacks that infer whether a sample was included in the training of the model, becomes effective by moving the sample to an outlier. However, the attacks can be detected because inference accuracy deteriorates due to poison samples. 
In this paper, we discuss a \textit{backdoor-assisted membership inference attack}, a novel membership inference attack based on backdoors that return the adversary's expected output for a triggered sample. 
We found three crucial insights through experiments with an academic benchmark dataset. We first demonstrate that the backdoor-assisted membership inference attack is unsuccessful. 
Second, when we analyzed loss distributions to understand the reason for the unsuccessful results, we found that backdoors cannot separate loss distributions of training and non-training samples. In other words, backdoors cannot affect the distribution of clean samples. Third, we also show that poison and triggered samples activate neurons of different distributions. 
Specifically, backdoors make any clean sample an inlier, contrary to poisoning samples.
As a result, we confirm that backdoors cannot assist membership inference. 
%我々がその原因を究明するためにloss distribution for inference を解析したところ、バックドアではthe adversary's expected output を得ることができる入力において、独立したloss 分布を作ってしまうことを確認した。各ニューロンの活性化状態においても、ポイズニングとバックドアには明確な違いを得ていることを確認している。このため、バックドア攻撃ではメンバーシップ推定攻撃をassist が難しいと確認している。
% 　次に既存のバックドア検知ツールを回避できるimperceptible backdoor attacks を用いることで、攻撃成功率に影響するか確認する。 extensive experiments を通じて調査したところ, まず既存攻撃と比べて同等程度の攻撃成功率を達成できることを確認した。  このとき, バックドア支援型メンバーシップ推定攻撃ではメンバーと非メンバーの分布の乖離がないまま攻撃が成功していることを確認している。次に、検地を回避できるバックドア攻撃を用いることで、攻撃成功率が低下することを確認した。検知回避の観点にはトリガーとlatent representation の二つがあるが、とくに後者が攻撃成功率を低下させる。
\end{abstract}


\section{Introduction} 
\label{sec:introduction}

Membership inference attacks are currently used for evaluating privacy leakage in various machine learning models~\cite{conti2022label-onlymembership,li2022auditingmembershipinference,ye2022enchancedmembership}. 
In a membership inference attack~\cite{ShokriSSS17}, an adversary infers whether a sample was utilized for training a machine learning model. 

In recent years, Tramer et al.~\cite{tramer2022truth} proposed an advanced attack, called poisoning-assisted membership inference attack, for amplifying privacy leakage by injecting poison samples into a dataset. 
The drawback of the attack is to deteriorate the inference accuracy of the victim model injected poison samples. 
Consequently, the owner of the victim model can detect the underlying poison samples in any kind of poisoning attack~\cite {tian2022acomprehensivesurvey}. 
Namely, the poisoning-assisted membership inference attack can be prevented by detecting poison samples; thus, it may be less severe than expected. 

The above limitation leads us to a membership inference attack utilizing a backdoor attack, i.e., \textit{a backdoor-assisted membership inference attack}. 
Backdoor attacks~\cite{Gu2019badnets} are stealthier than poisoning attacks because they manipulate the output of only triggered samples and maintain test accuracy~\cite{tian2022acomprehensivesurvey}. 
There are also advanced attacks, called imperceptible backdoors~\cite{li2020invisible,ning2021invisible,doan2021lira,zhong2020backdoorembedding,tan2020bypassing,Tang2021demon,Zhao2022defeat,zhong2022imperceptible}, that bypass existing backdoor detection tools~\cite{wang2019neuralcleanse,Chen2019detecting,chou2020sentinet,gao2019strip,liu2019abs,tran2018spectral}. 

In this paper, we take the first step for answering two following key questions on backdoor-assisted membership attacks. 
(1) \textit{Is a backdoor-assisted membership inference attack feasible?}
(2) \textit{Do backdoors impact the loss distributions of a victim model?}

The above questions are non-trivial.
For the first question, it is unclear whether a backdoor-assisted membership inference attack works because the backdoor attacks maintain inference accuracy. 
The key idea of the existing poisoning-assisted membership inference attack~\cite{tramer2022truth} is to make the target sample an outlier by deteriorating accuracy with poison samples. 
In contrast, backdoors may not make the target sample an outlier because they maintain accuracy. 
For this reason, the backdoor-assisted membership inference attack is significantly different from the existing attack. 

Next, for the second question, the difference between loss distributions of training and non-training data is also essential for amplifying the privacy leakage. 
%How different between the loss distributions on backdoor-assisted membership inference attacks compared to poisoning-assisted attacks? 
It is known that a membership inference attack is statistical testing with loss distribution~\cite{carlini2022firstprinciples}, and then poison samples can boost membership inference attacks by separating loss distributions between training and non-training samples~\cite{tramer2022truth}.
However, it is unclear whether the backdoors can boost membership inference attacks because the difference between the loss distributions of a backdoor-assisted attack and a poisoning-assisted attack has never been discussed. 
%for the same reason as poison samples.

We found three crucial insights through the experiments with a typical academic benchmark, i.e., the CIFAR-10 dataset. 
As the first insight, the backdoor-assisted membership inference attack is \textit{unsuccessful} as opposed to the poisoning-assisted membership inference attack. 
Specifically, a backdoor attack amplifies only few attack success rates of the membership inference.

Next, we analyze loss distributions and neuron activations of the victim models to understand the above phenomenon deeply. 
Then, as the second insight, we found that backdoors cannot separate loss distributions of training and non-training samples.
We also demonstrate, as the third insight, that the backdoor-assisted membership inference attack makes a target sample an inlier in the distribution of activated neurons. 
In contrast, the poisoning-assisted membership inference attack makes it an outlier. 
We thus believe that backdoors \textit{do not} assist in membership inference attacks. 

To sum up, we found the following crucial insights in this paper: 
\begin{itemize}
    \item Backdoor-assisted membership inference attacks are unsuccessful.
    \item Backdoors do not separate loss distributions of training and non-training samples.
    \item Backdoor-assisted membership inference attacks make a target sample an inlier, while poisoning-assisted membership inference attacks make it an outlier. 
\end{itemize}

\section{Related Work} \label{sec:related_works}

%本節では関連研究を紹介する. 
%（定式化まではしない）

In this section, we describe related works of backdoor attacks and privacy violations assisted by poisoning attacks.

\subsection{Backdoor Attacks} %矢内
\label{sec:related_backdoors}

Backdoor attacks~\cite{Gu2019badnets,Liu2018trojaning} are a kind of attack whereby an adversary trains a model such that he/she obtains the expected output for only triggers. 
Recent backdoor attacks~\cite{li2020invisible,ning2021invisible,zhong2020backdoorembedding,xue2021backdoors,saha2020hiddentrigger,li2021hiddenbackdoors} can bypass detection tools~\cite{chou2020sentinet,wang2019neuralcleanse,tran2018spectral,gao2019strip,Chen2019detecting,liu2019abs}, and thus existence of backdoors is imperceptible. (We call them imperceptible backdoor attacks for the sake of convenience.) 

There are three approaches for constructingimperceptible backdoor attacks. 
The first approach~\cite{li2020invisible,ning2021invisible,doan2021lira,zhong2020backdoorembedding} is based on trigger generation that is visually imperceptible for humans, referred to as the trigger method. 
The second approach~\cite{tan2020bypassing,Tang2021demon} is based on latent representations whose distributions are close between clean inputs and triggers, referred to as the latent-representation method. 
The third approach~\cite{Zhao2022defeat,zhong2022imperceptible} is unified attacks~\cite{Zhao2022defeat,zhong2022imperceptible} for the above two approaches, referred to as the unified method. 
We evaluate backdoor-assisted membership inference attacks based on the above three imperceptible backdoor attacks as well as the original backdoor attack~\cite{Gu2019badnets}. 

In recent years, backdoor attacks have been discussed in real-world applications such as natural language processing~\cite{li2021hiddenbackdoors,qi2021hiddenkiller} and face authentication~\cite{xue2021backdoors}. 
Combining our attack with these works makes privacy violations in real-world applications possible.

\subsection{Privacy Violations Assisted by Poisoning
%ポイズニング支援型メンバーシップ推論攻撃
} %芦澤

There are existing works on privacy violations based on poisoning attacks~\cite{mahloujifar2022property,tramer2022truth,hidano2018modelinversion,wang2022poisoning-assisted,chen2022amplifying}. 
The first work~\cite{hidano2018modelinversion} was in simple models such as support vector machines. 
Whereas several papers~\cite{mahloujifar2022property,wang2022poisoning-assisted} discussed property inference attacks~\cite{ganju2018propertyinference} that infer properties of a training dataset, Tramer et al.~\cite{tramer2022truth} discussed a membership inference attack, attribute inference~\cite{fredrikson2014privacy,fredrikson2015modelinversion,yeom2018privacyrisk}, and data extraction~\cite{carlini2019secretsharer,carlini2021extracting,henderson2018ethical}. 
Nevertheless, the above works did not discuss backdoor attacks. 
In other words, the above works succeeded in membership inference attacks by sacrificing accuracy~\cite{tramer2022truth}. 
%In this section, we describe related works of privacy violations assisted by poisoning attacks and backdoor attacks. 

The closest work to ours is by Chen et al.~\cite{chen2022amplifying}. 
They evaluated a membership inference attack with clean-label poisoning~\cite{shafahi2018poisonfrogs,zhu2019transferlablecleanlabel}, whose labels remain unchanged and samples are visually indistinguishable from clean samples. 
However, in their attack, the distance between clean and poison samples in latent representations still becomes far from each other to maximize the influence of the target. 
As described in the previous subsection, we discuss not only the original backdoor attack~\cite{Gu2019badnets}, which is regardless of the distance between clean and poison samples, but also the imperceptible backdoor attacks whose distance between clean and poison samples in latent representations is close to each other. 
Especially, the latter backdoor attacks extremely differ from Chen et al.'s work. 
%We discuss backdoor attacks whose distance between clean and poison samples in latent representations is close to each other.
%We discuss backdoor も
%backdoorにはうえのような違いがある (poisoningと)
%amplifyingには使われていない上記のようなbackdoorについても

Although several works~\cite{li2022untargetedbackdoor,hu2022membershipinference} combine backdoors with membership inference, they are close to watermarking~\cite{WatermarkNN} to check if a model is backdoored. 
Namely, these works are quite different from privacy violations, i.e., our leading problem, because they infer backdoors embedded by a model owner.

\section{Backdoor-Assisted Membership Inference Attack} %矢内

We describe a backdoor-assisted membership inference attack as the problem setting of this paper below. 
We first define an attack formally and its metrics. 
We then describe the key questions in detail.

\subsection{Formalization} \label{sec:formalization}

The attacks in this paper are defined as a game between an adversary $\mathcal{A}$ and a challenger $\mathcal{C}$. 
We first denote by $\mathcal{X}$ a set of data samples, by $\mathcal{Y}$ a set of labels, by $\mathcal{D} = \mathcal{X} \times \mathcal{Y}$ a set of datasets, and by $\mathcal{M}$ a set of machine learning models. 
Then, a machine learning model $M \in \mathcal{M}$ is defined as a mapping function $M: \mathcal{X} \rightarrow \mathcal{Y}$ and a training algorithm, i.e., a loss function, is defined as a mapping function $L_M: \mathcal{D} \rightarrow \mathcal{M} $. 

The game is defined below. 
$\mathcal{A}$ then interacts with the final trained model.
$\mathcal{A}$ is in the black-box setting that he/she sends queries to the model $M$ and obtains nothing more than the outputs. 
Here, $\mathcal{A}$ can only provide statically poison samples. 

\begin{enumerate}
    \item $\mathcal{C}$ chooses a clean dataset $D\subseteq \mathcal{D}$. 
    
    \item $\mathcal{C}$ chooses a bit $b\leftarrow \{0,1\}$. 
    If $b=1$, $\mathcal{C}$ chooses a sample $z\in D$; otherwise, $\mathcal{C}$ chooses $z\in \mathcal{D}\backslash D$. 
    
    \item Given $z$, $\mathcal{A}$ chooses a poisoning dataset $D_p \subset \mathcal{D} $ consisting of $n$ samples, and send it to $\mathcal{C}$. 
    
    \item $\mathcal{C}$ trains $M$ by $M=L_M(D^*)$ with the entire dataset $D^* = D\cup D_p$. 
    \textcolor{red}{In doing so, assuming $M^*=L_{M^*}(D)$ for any model $M^* \in \mathcal{M}$, $M^*$ and $M$ achieve the following relations: 
    (1) for any $(x,y) \in D$, $M(x)=M^*(x)$ holds; 
    and, (2) for any $(x^*,y^*) \in D_p$, $M(x^*) = y^*$.
    } 
    
    \item $\mathcal{A}$ sends samples $x_1, \cdots, x_q \in \mathcal{X}$ to $M$ and obtains $y_1 = M(x_1), \cdots, y_q = M(x_q)$. 
     
    \item $\mathcal{A}$ returns a bit $b'\in \{0,1 \}$. 
    If $b=b'$ holds, $\mathcal{A}$ wins the game. 
\end{enumerate}

In the above game-based definition, the sentences with red color differ from the existing attack by Tramer et al.~\cite{tramer2022truth}. 
While an adversary in the existing attack does not require a model $M$ anything other than learning a dataset $D_p$, our adversary requires a model $M$ to misinfer only a sample $x^*$ in $D_p$ as his/her expected output $y^*$. 
It is the requirement of backdoor attacks~\cite{Gu2019badnets}.

\subsection{Evaluation Metrics}

We adopt the following evaluation metrics in this paper. 
%We adopt attack success rate~\cite{ShokriSSS17} and area-under-the-ROC-curve (AUC)~\cite{song2019auditing} as evaluation metrics of the attacks. 

\textbf{Membership-inference-attack success rates (MIA-SR)~\cite{ShokriSSS17}:}
For the number n of execution times of the game described in the previous section and the number $a$ of times that $\mathcal{A}$ wins the game, it is defined as $a/n$. 

\textbf{Membership-inference-attack AUC (MIA-AUC)~\cite{song2019auditing}:}
MIA-AUC is represented as an area under the ROC curve. 
The ROC curve is a two-dimensional curve defined by true positive rates (TPR) and false positive rates (FPR). 
They mean that positive and negative values are accurately estimated as members and non-members. 

In addition to the above metrics, we introduce metrics for backdoor attacks~\cite{Gu2019badnets}, i.e., test accuracy and backdoor identification rates\footnote{It is originally defined as the attack success rate in~\cite{Gu2019badnets}, but we say backdoor identification rate for convenience.}. 
For a model $M = L_M(D^*)$ with the entire dataset $D^* = D \cup D_p$ and another model $M^* = L_{M^*}(D)$ with only a clean dataset $D$, they are defined as follows:

\textbf{Test Accuracy (TA):} 
For any pair $(x,y) \in D \subseteq \mathcal{D}\backslash D_p$ of a clean sample and its label, it is defined as a ratio such that $y = M (x)$ holds, where $M(x) = M^*(x)$ holds to maintain the original inference by $D$ possibly. 
Intuitively, accuracy is necessary for stealthy compared to conventional poisoning attacks.

\textbf{Backdoor identification rates (BIR):} 
For any pair $(x^*,y^*) \in D_p$ of a poison sample and its label, it is defined as a ratio such that $M(x^*) = y^*$ holds, where $M(x^*) \neq M^*(x^*)$ may hold. 
It means that an adversary $\mathcal{A}$ can certainly exploit backdoors embedded in $M$.

\subsection{Key Questions} \label{sec:key_question}

We have two key questions about backdoor-assisted membership inference attacks. 
First, we discuss the impact of the difference from the existing works~\cite{tramer2022truth,chen2022amplifying}, i.e., the sentences with red color, on the attacks. 
We then evaluate MIA-SR and MIA-AUC with respect to TA and BIR.
Second, we discuss loss distributions between training and non-training samples for each attack.

Table~\ref{tab:list_of_attacks} summarizes the primary differences from the existing works~\cite{tramer2022truth,chen2022amplifying}.

\begin{table}[t!]
    %\centering
    \caption{Property of Poisoning-/Backdoor-Assisted Attacks}
    \label{tab:list_of_attacks}
    We investigate and discuss the following attack methods. 
    For the column of ``Accuracy," the checkmark means that backdoor attacks keep the accuracy. 
    For the columns of ``Trigger" and ``Latent Representation," the checkmarks mean imperceptible backdoors in each context. %\\ \vspace{1zw}
    \begin{tabular}{cc|c|c|c}
    \hline
         & Method & Accuracy & Trigger & Latent Representation  \\ \hline \hline
          & Truth Serum~\cite{tramer2022truth} & & &  \\ \hline 
        & Chen et al.~\cite{chen2022amplifying} &  & \checkmark &  \\ \hline
        Our & BadNets~\cite{Gu2019badnets} & \checkmark &  &  \\ \hline
        Our & TaCT~\cite{Tang2021demon} & \checkmark &   & \checkmark \\ \hline
        Our &    LIRA~\cite{doan2021lira} & \checkmark  & \checkmark &  \\ \hline
        Our & IBD~\cite{zhong2022imperceptible}      &  \checkmark & \checkmark & \checkmark \\ \hline
    \end{tabular}
\end{table}

\section{Experiment} %後藤

We conduct extensive experiments with our backdoor-assisted membership inference attacks. 
As described in the previous section, our goal is to discuss the impact of backdoors on MIA-SR and MIA-AUC by comparing them with the existing attack by Tramer et al.~\cite{tramer2022truth}.

We follow the targeted attack setting by Tramer et al.~\cite{tramer2022truth}, where an adversary targets a specific example. 
The attacks were implemented with PyTorch.
The source code for the attacks is published in GitHub\footnote{URL will be updated in the later versions.}.

\subsection{Setting}

\subsubsection{Model and Baseline}
We also trained six ResNet18 models: 
(1) model attacked with neither poisoning nor backdoor attacks, referred to as Clean-Only model; 
(2) model referred to as Truth Serum~\cite{tramer2022truth} as a baseline attacked with $250 \times r$ poison samples in $D_p$ for $r \in \{1,2,4,8,16\}$;
and (3)-(6) models backdoor-attacked with $250 \times r$ triggered samples in $D_p$ based on BadNets~\cite{Gu2019badnets}, TaCT~\cite{Tang2021demon} LIRA~\cite{doan2021lira}, andIBD~\cite{zhong2022imperceptible}, respectively. 
We refer to each model of (3)-(6) as BadNets, Tact, LIRA, and IBD.

\subsubsection{Dataset}
We utilize the CIFAR-10 dataset for the experiment. 
The $250$ samples in $D_p$ for poisoning or trigger are extracted from 50,000 training samples of the CIFAR-10 dataset.
Also, the 50,000 training samples are divided into two groups: the training dataset $D$ and the test dataset $\mathcal{D}\backslash D$ of the victim model.
The victim model learns the entire dataset $D^* = D\cup D_p$.
Here, for the clean model, 25,000 samples are randomly chosen from 50,000 training samples as $D$. 
In the same way, the dataset for the shadow model is prepared.

\subsubsection{Membership Inference Attack Method}

We implemented the membership inference by Carlini et al.~\cite{carlini2022firstprinciples}. 
Their attack needs shadow models to mimic the data distribution of a victim model $M$; hence, we prepare twenty models. 
We then choose the victim model $M$ from the twenty models, and the remaining models are utilized for shadow models to conduct a full leave-one-out cross-validation. 
We measure MIA-SR and MIA-AUC on these six models and then evaluate their results.

\subsection{Results} \label{sec:exp_results}

\begin{figure*}[tb]
\begin{subfigmatrix}{3}
\subfigure[Clean-Only]{\includegraphics[width=.30\textwidth]{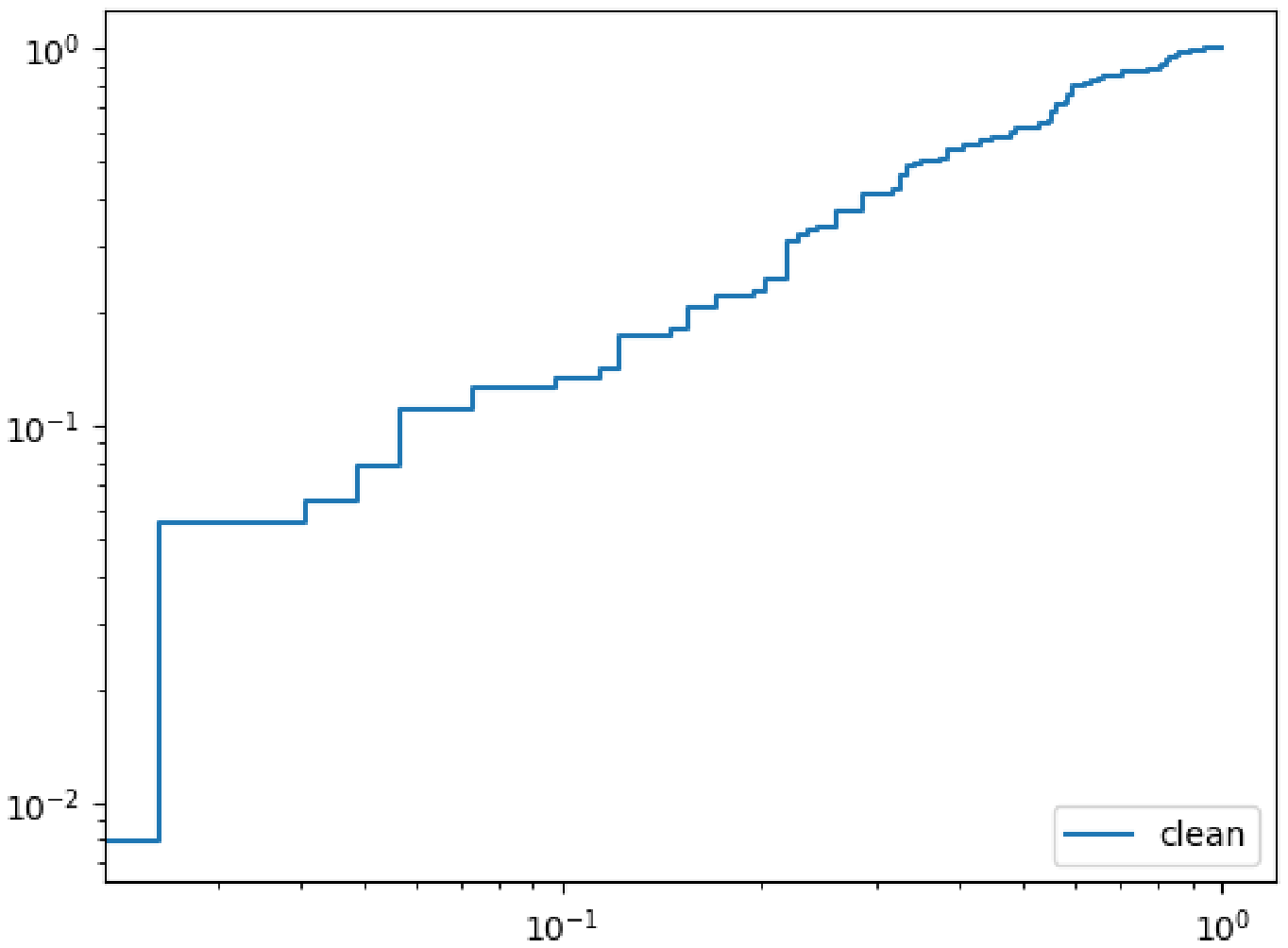}}
\subfigure[Truth Serum]{\includegraphics[width=.30\textwidth]{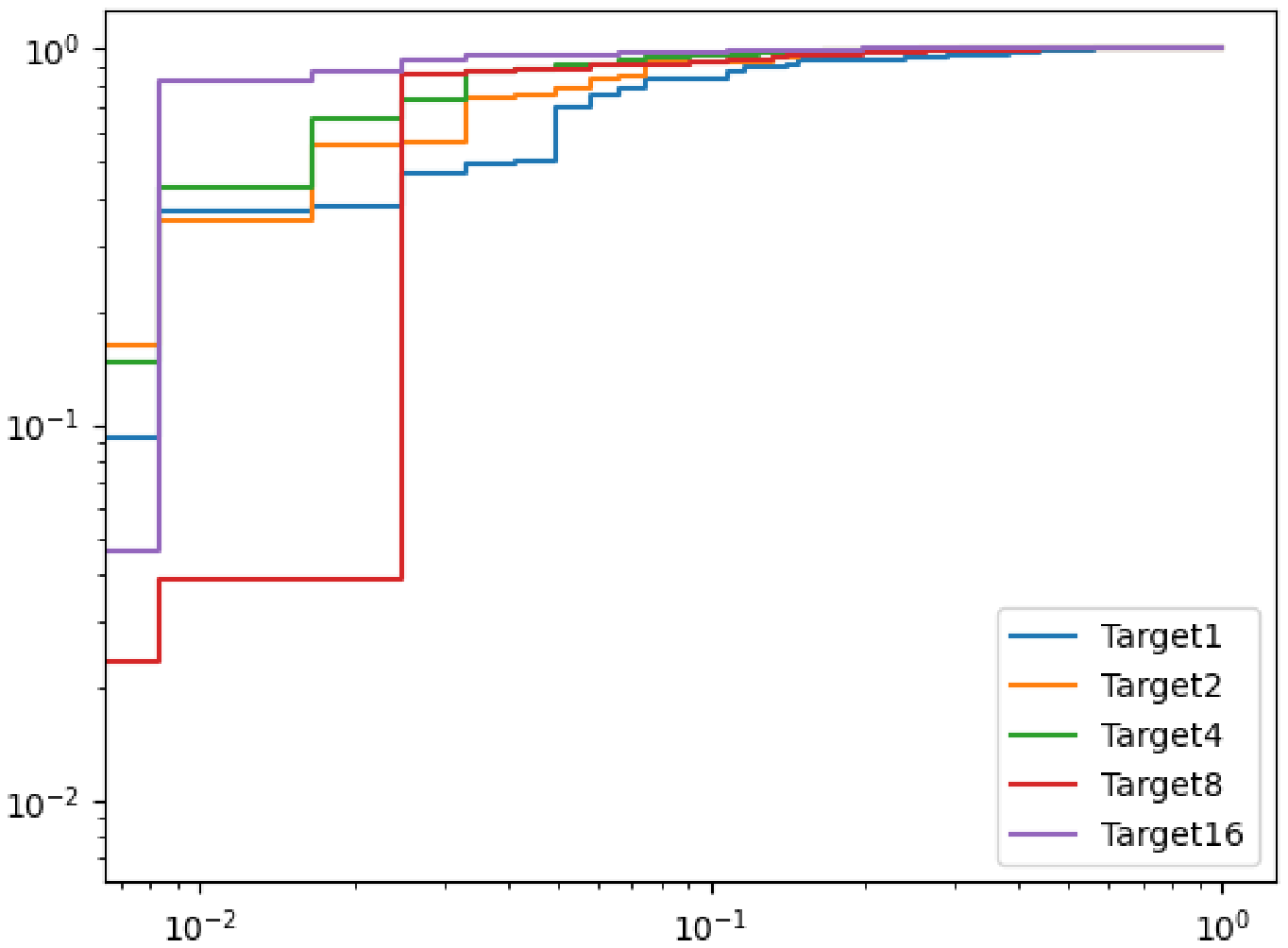}}
\subfigure[BadNets]{\includegraphics[width=.30\textwidth]{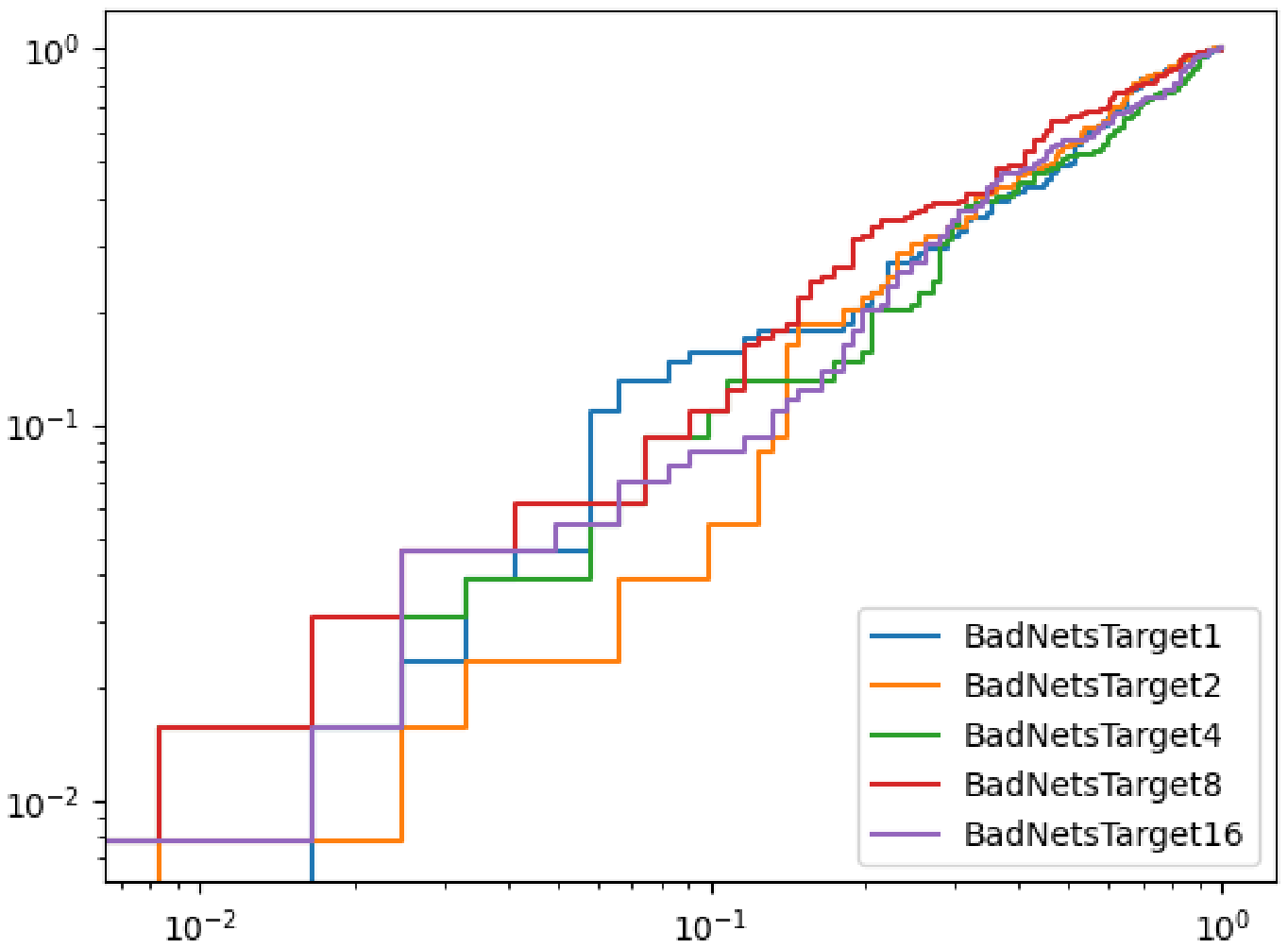}}
\subfigure[TaCT]{\includegraphics[width=.30\textwidth]{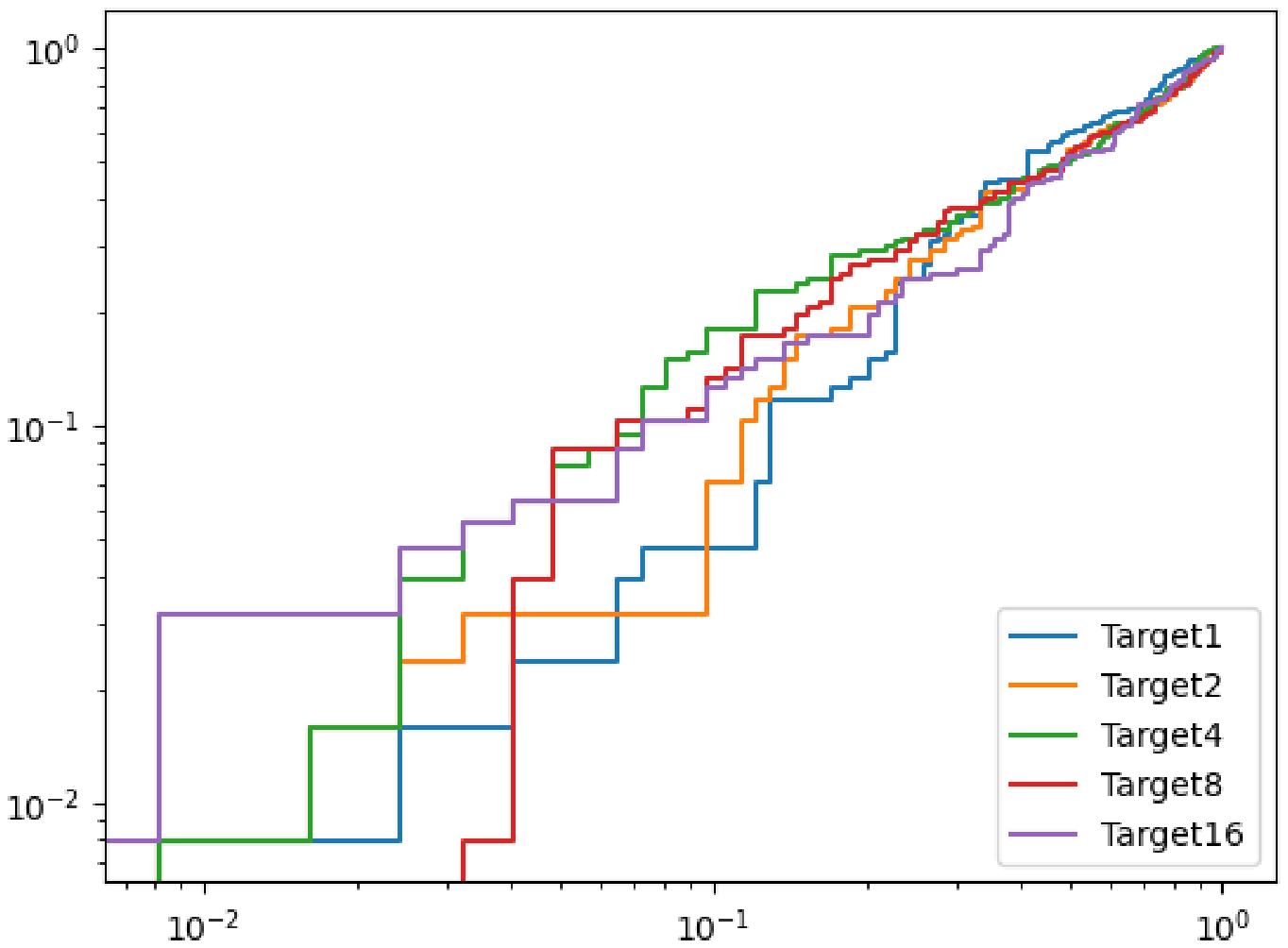}}
\subfigure[LIRA]{\includegraphics[width=.30\textwidth]{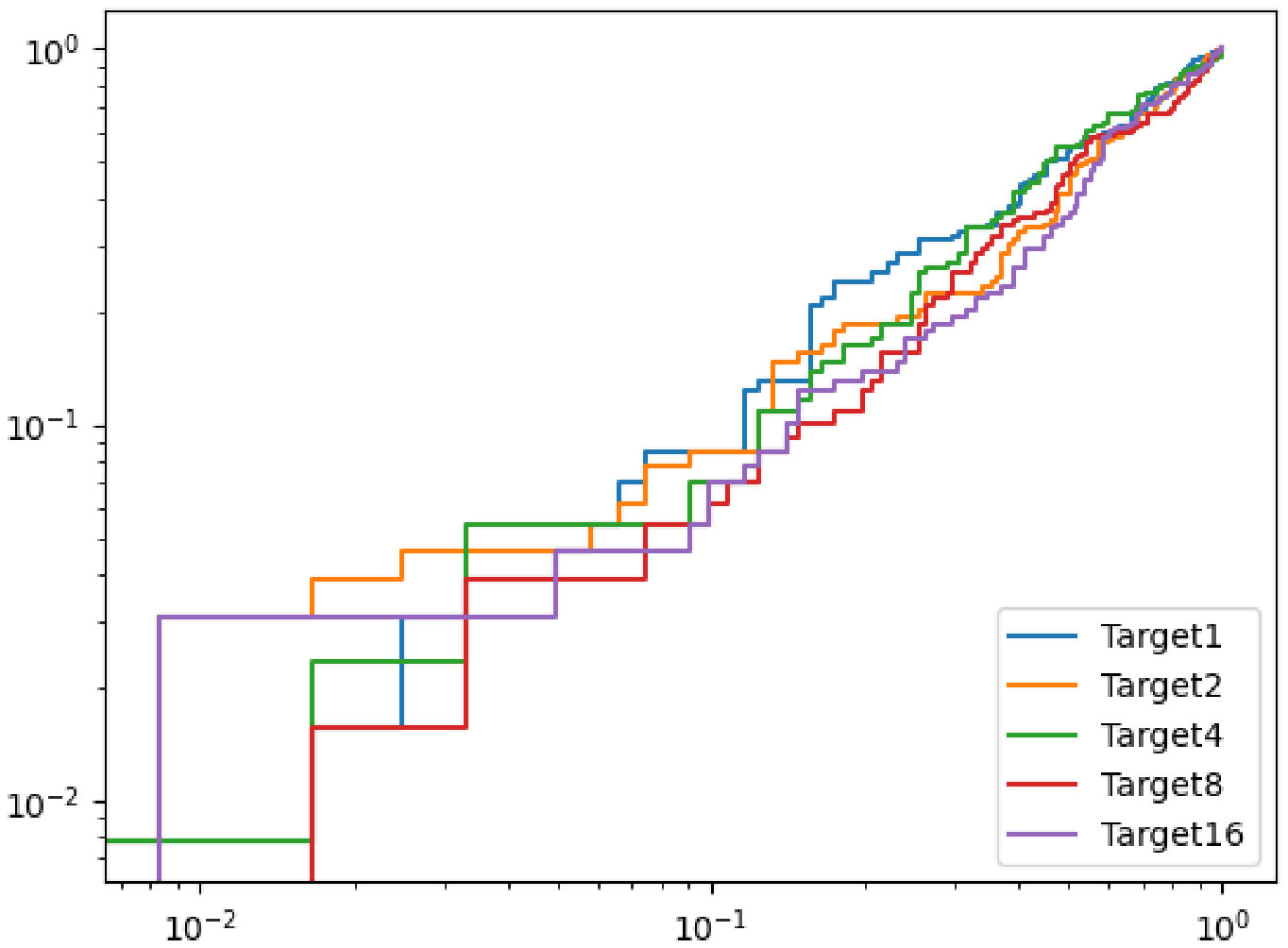}}
\subfigure[IBD]{\includegraphics[width=.30\textwidth]{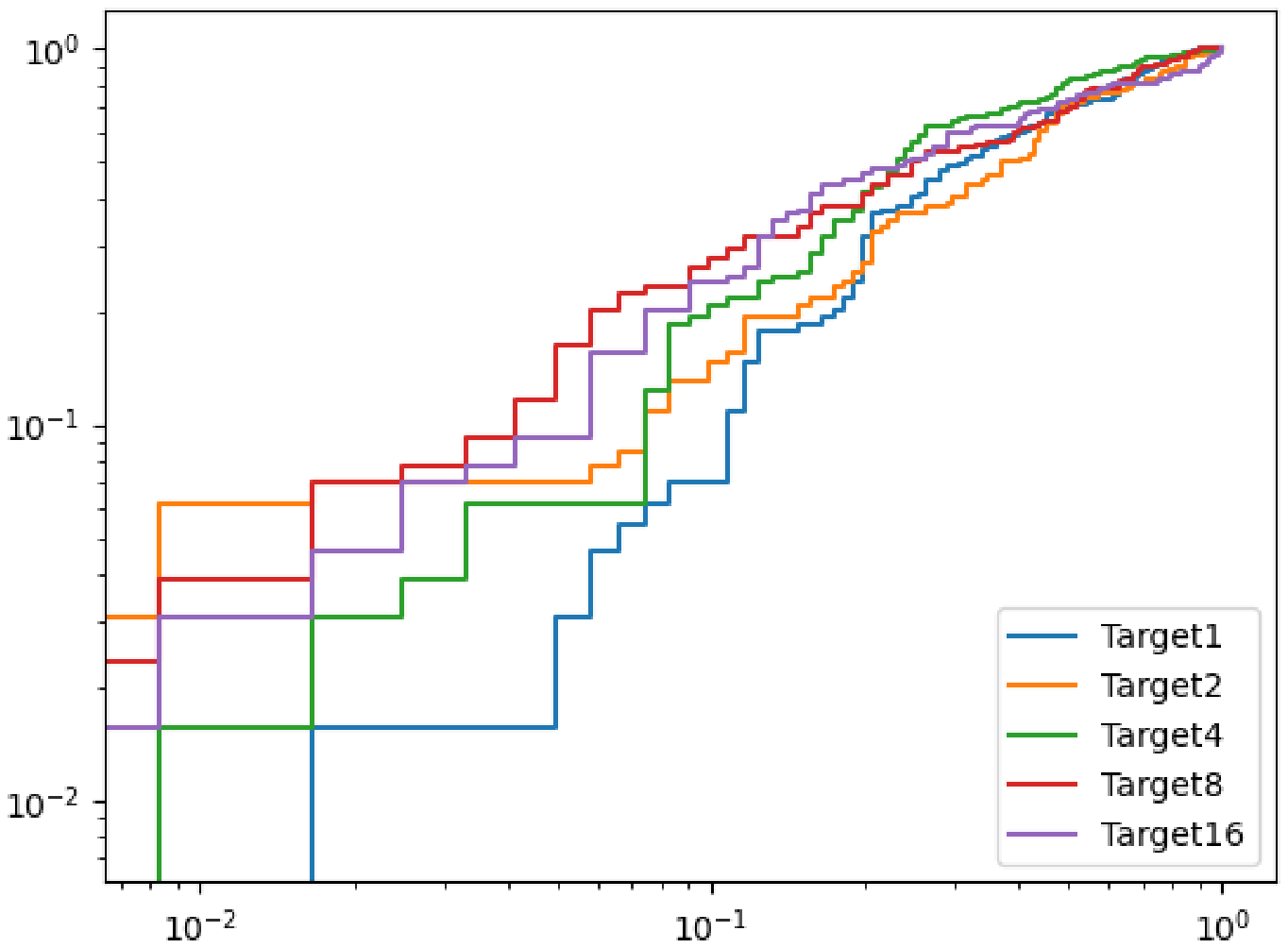}}
\end{subfigmatrix}
\caption{MIA-ROC for Each Attack}
\label{fig:roc}
Poisoning attacks boost membership inference attacks with 250 target poisoning data. 
In proportion to the number of poisoning samples, the true positive rates become higher while false positive rates become lower. 
Backdoor attacks have less influence on the true positive and false positive rates of membership inference attacks.
\end{figure*}

Table~\ref{tab:MIA_result_clean} shows the result of the membership inference attack against the Clean-Only model, which is identical to the baseline. 
Fig.~\ref{fig:roc} and Table~\ref{tab:MIA_result_backdoor} show the results of each attack.

According to the table, BadNets and TaCT, which need a few triggers for backdoors, keep high test accuracy, unlike Truth Serum.
However, when we compare BadNets with TaCT, both MIA-SR and MIA-AUC deteriorate for any number of samples.
It indicates that MIA-SR and MIA-AUC deteriorate due to triggers generated in imperceptible approaches.
On the other hand, MIA-AUC of Truth Serum increases more than or equal to 0.32 compared to the Clean-Only model.

We also explain why TA and BIR for MIA-SR and MIA-AUC are low below. 
LIRA and IBD need the same number of triggers as clean samples for backdoors.
In this experiment, we used at most 4000 triggers for LIRA and IBD, despite training with 25000 clean samples.
That is, TA and BIR for LIRA and IBD became low compared to BadNets and TaCT. 
%why the test accuracy and the backdoor identification rates of LIRA and IBD are low compared to other backdoors.

\begin{table}[tbp]
 \caption{Result of Membership Inference Attack on Clean-Only Model}
 \label{tab:MIA_result_clean}
 The Clean-Only model infers classes of non-training data with high test accuracy. 
 When MIA-SR and MIA-AUC are close to 50\%, it is random. 
 \centering %\\ \vspace{1zw}
  %(by First Principle.)
% \centering
 \begin{tabular}{|c|ccc|}
    %\centering
     \hline
     %小数点2以下切り捨て
     & TA & MIA-SR [\%] & MIA-AUC \\ % Clean
     \hline \hline
     Clean-only & 90.94 & 58.30 & 0.60 \\
     \hline
    \end{tabular}
\end{table}

\begin{table*}[tbp]
\caption{Results of Each Attack}
\label{tab:MIA_result_backdoor}
MIA-SR becomes higher for Truth Serum, but lower by each backdoor attacks. 
TA, BIR, and MIA-SR are the percentage of leave-one-out cross-validation with 20 shadow models.
\centering \\
\SetFigLayout{2}{1}
 %小数点3以下切り捨て
 \subfigure[]{
    \begin{tabular}{|c||cccc||cccc|}
     \hline
     & \multicolumn{4}{|c||}{$r=$1} 
     & \multicolumn{4}{|c|}{$r=$2} \\
     
     & TA & BIR & MIA-SR & MIA-AUC 
     & TA & BIR & MIA-SR & MIA-AUC \\
     \hline \hline
     
     Truth Serum
     & 85.94 & - & 86.08 & 0.92 
     & 85.85 & - & 91.34 & 0.96 \\
     BadNets
     & 90.85 & 94.72 & 56.68 & 0.57 
     & 90.83 & 94.45 & 55.28 & 0.57 \\
     TaCT
     & 90.84 & 80.48 & 51.86 & 0.52 
     & 90.86 & 80.73 & 51.48 & 0.52 \\
     LIRA
     & 43.49 & 16.45 & 50.12 & 0.50 
     & 49.62 & 18.99 & 49.68 & 0.49 \\
     IBD
     & 71.19 & 15.75 & 55.86 & 0.60 
     & 74.06 & 14.32 & 55.90 & 0.62 \\
     \hline
    \end{tabular}
    }
    \hfill
    \subfigure[]{
    \begin{tabular}{|c||cccc||cccc||cccc|}
     \hline
     & \multicolumn{4}{|c||}{$r=$4} 
     & \multicolumn{4}{|c||}{$r=$8} 
     & \multicolumn{4}{|c|}{$r=$16} \\
     
     & TA & BIR & MIA-SR & MIA-AUC 
     & TA & BIR & MIA-SR & MIA-AUC 
     & TA & BIR & MIA-SR & MIA-AUC \\
     \hline \hline
     
     Truth Serum
     & 85.52 & - & 93.48 & 0.97 
     & 85.20 & - & 93.89 & 0.98 
     & 84.66 & - & 95.09 & 0.98 \\
     BadNets
     & 90.82 & 94.57 & 54.40 & 0.55 
     & 90.83 & 94.41 & 54.04 & 0.56 
     & 90.82 & 94.34 & 53.68 & 0.55 \\
     TaCT
     & 90.87 & 81.06 & 51.04 & 0.51 
     & 90.86 & 81.72 & 50.62 & 0.50 
     & 90.84 & 82.21 & 51.28 & 0.51 \\
     LIRA
     & 53.06 & 39.20 & 50.98 & 0.51 
     & 61.98 & 62.16 & 49.92 & 0.50 
     & 68.13 & 70.29 & 50.12 & 0.50 \\
     IBD
     & 75.00 & 13.55 & 54.04 & 0.62 
     & 75.90 & 12.93 & 53.84 & 0.64 
     & 77.05 & 12.49 & 53.28 & 0.64 \\
     \hline
    \end{tabular}
    }
\end{table*}

% clean only, PoisoningとTaCT, IJCAIのROC曲線

\section{Discussion} %後藤

In this section, we discuss why MIA-SR and MIA-AUC of the backdoor-assisted membership inference attack deteriorate from two standpoints. 
First, we analyze the impact of backdoors on loss distributions of a victim model because they are essential for MIA-SR in general~\cite{carlini2022firstprinciples}.
Second, we analyze the impact of backdoors on neuron activation to understand the internal parameters of the victim model in detail.

\subsection{Impact of Backdoors on Loss Distributions}

We discuss the impact of backdoors on loss distributions of a victim model for the backdoor-assisted membership inference attack. 
It is considered that a membership inference attack is a statistical testing with loss distribution~\cite{carlini2022firstprinciples}. 
Hence, we analyze the loss distribution for each model. 

\begin{figure*}[h!]
\begin{subfigmatrix}{5}
\subfigure[Truth Serum for $r=$ 2]{\includegraphics[width=.15\textwidth]{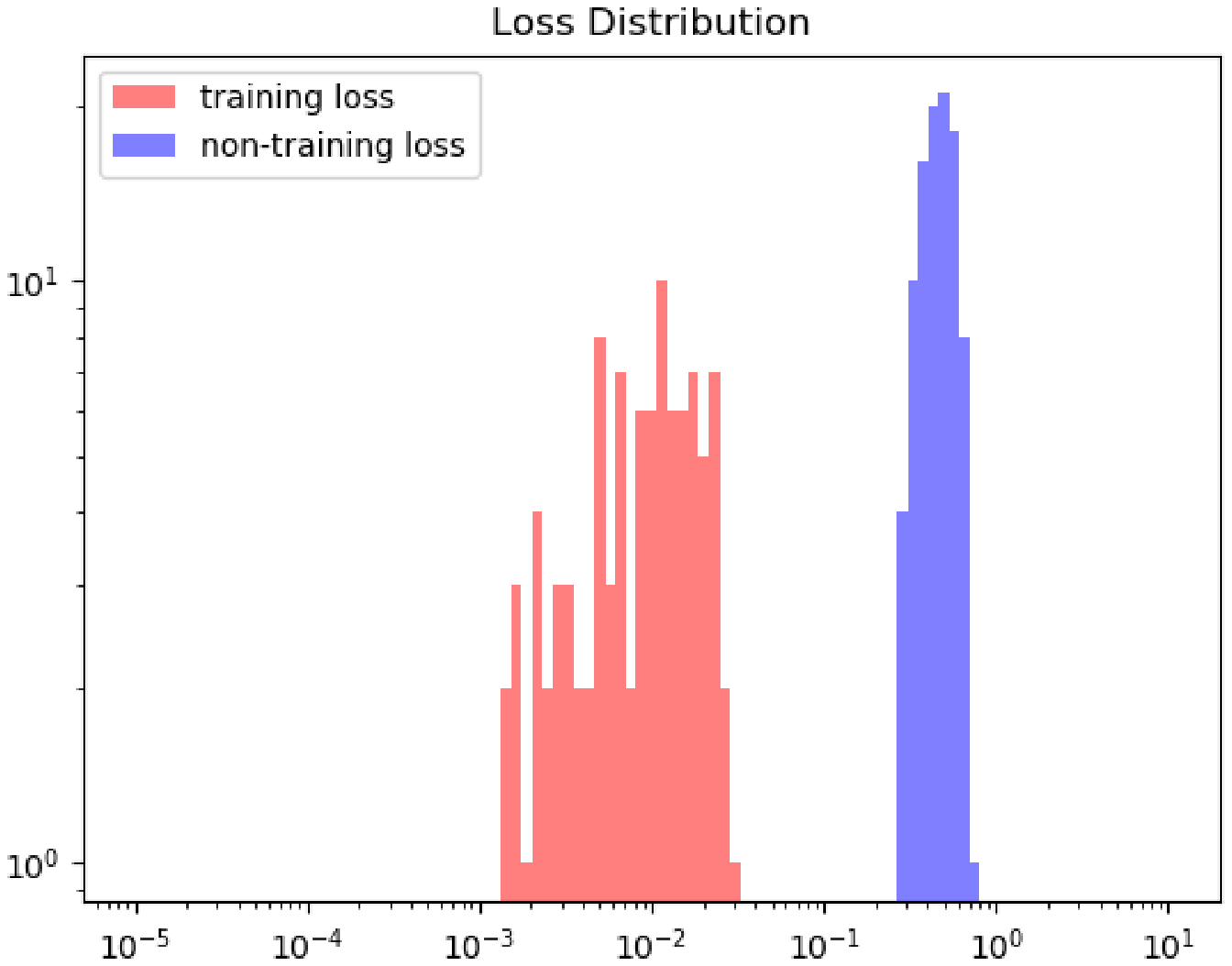}}
\subfigure[BadNets for $r=$ 2]{\includegraphics[width=.15\textwidth]{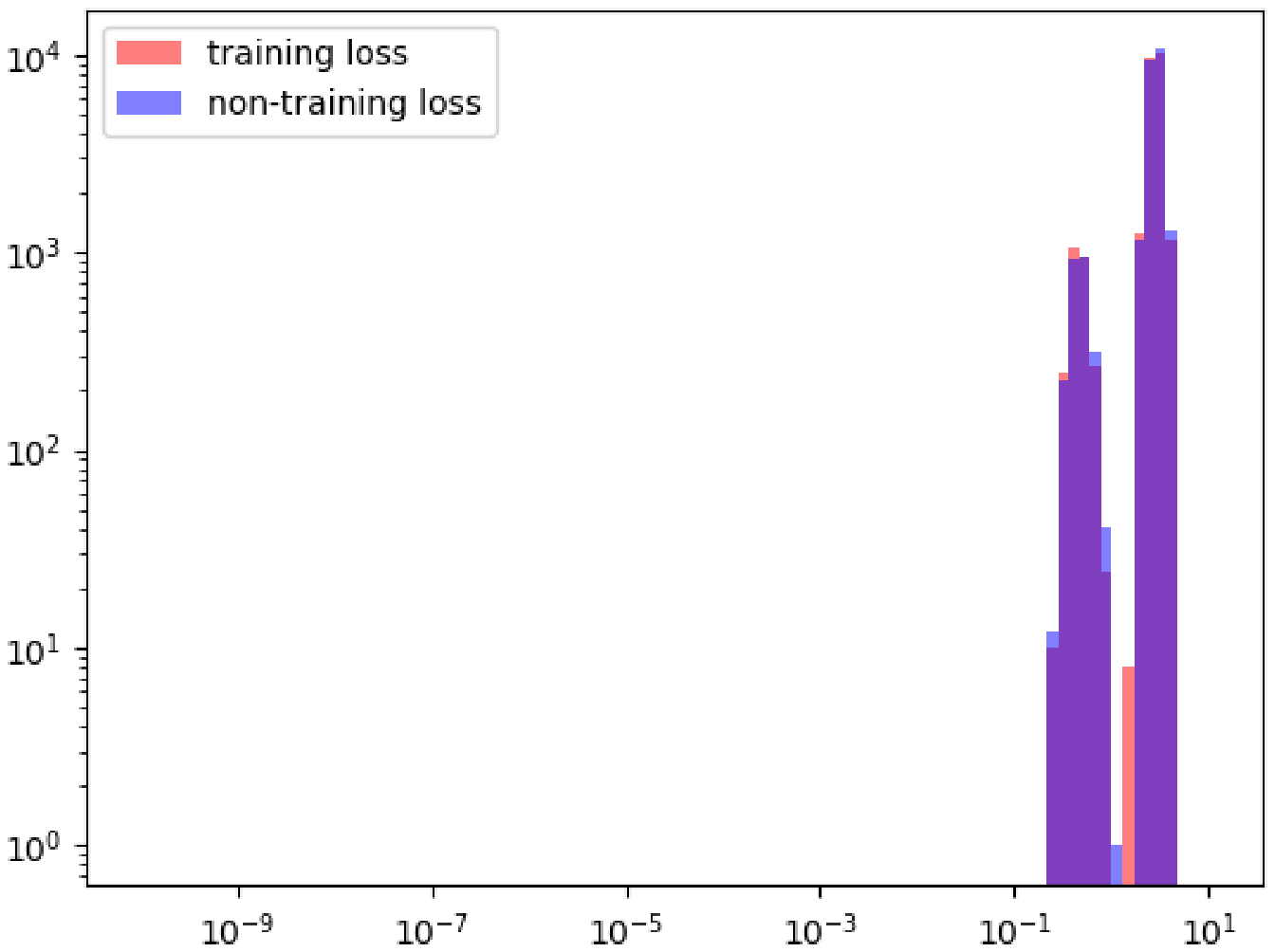}}
\subfigure[TaCT for $r=$ 2]{\includegraphics[width=.15\textwidth]{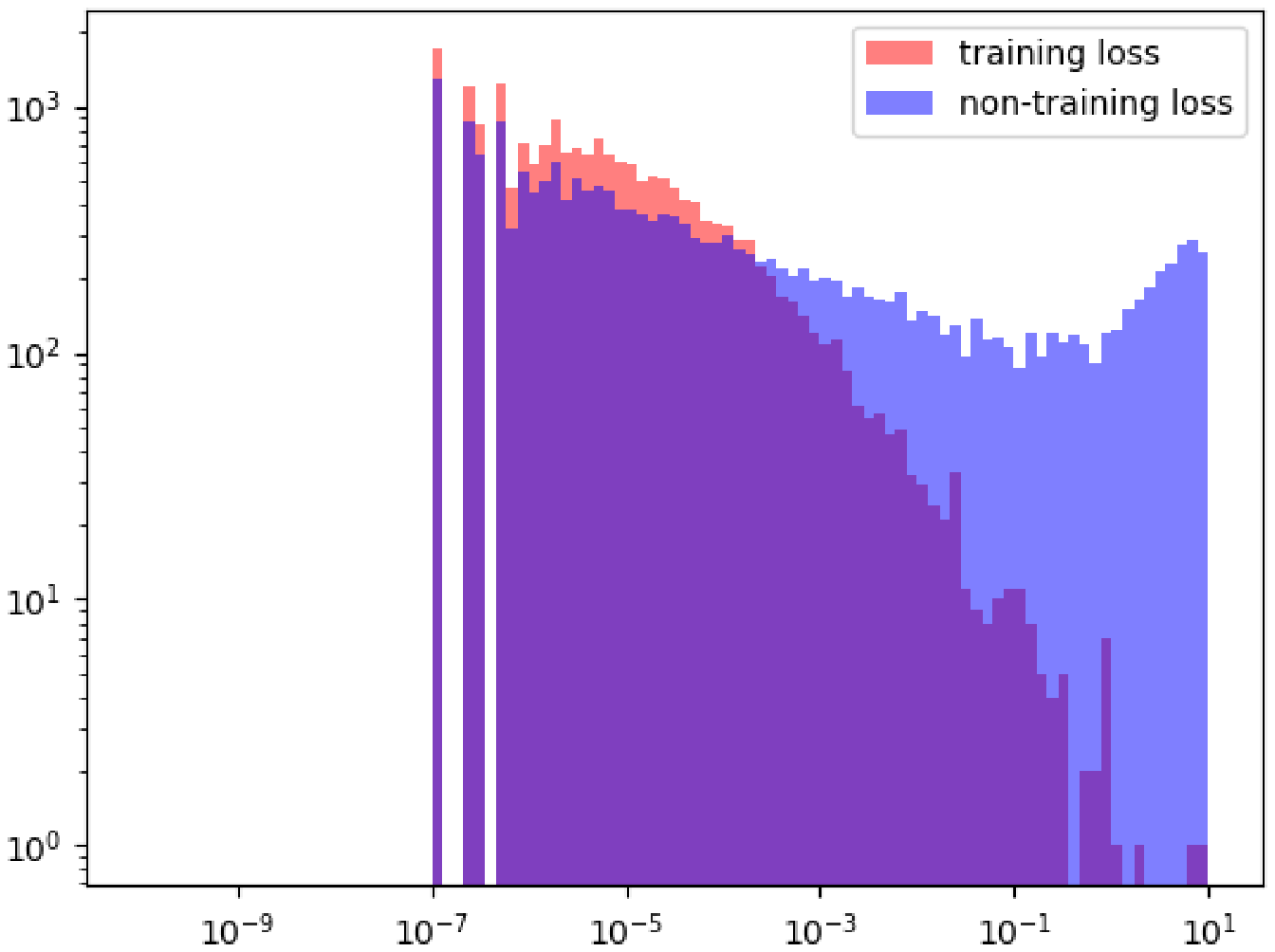}}
\subfigure[LIRA for $r=$ 2]{\includegraphics[width=.15\textwidth]{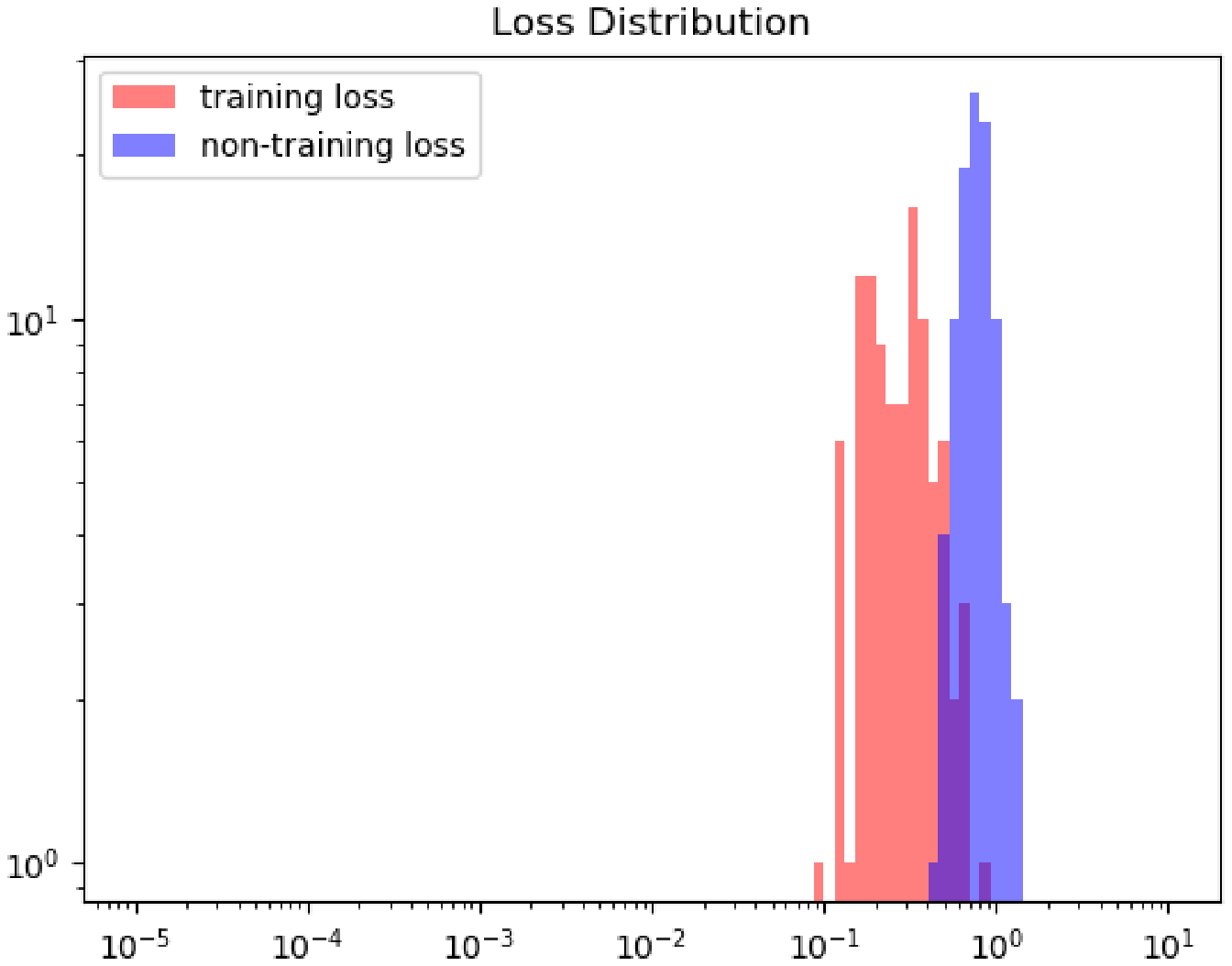}}
\subfigure[IBD for $r=$ 2]{\includegraphics[width=.15\textwidth]{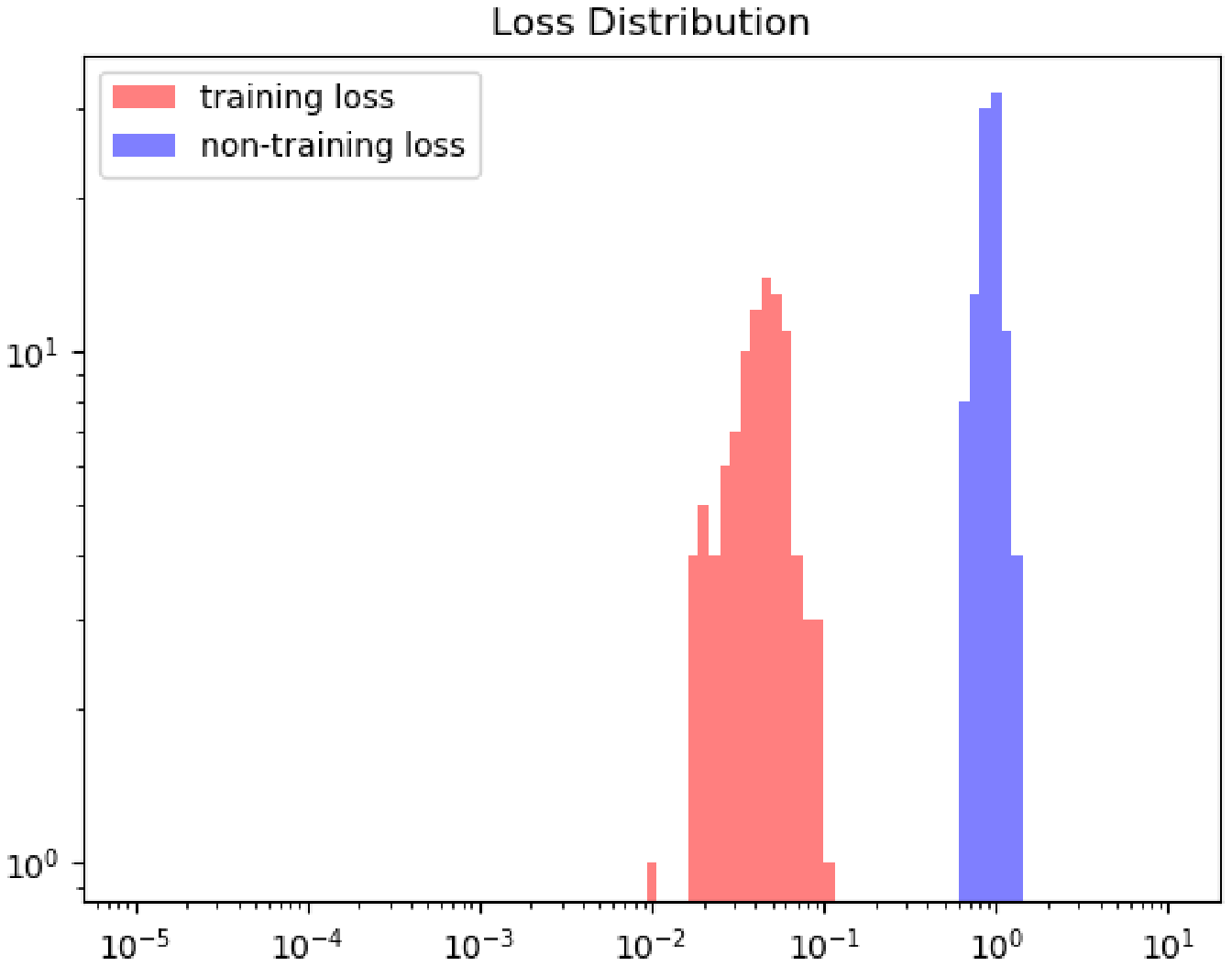}}
%%%%
\subfigure[Truth Serum for $r=$ 8]{\includegraphics[width=.15\textwidth]{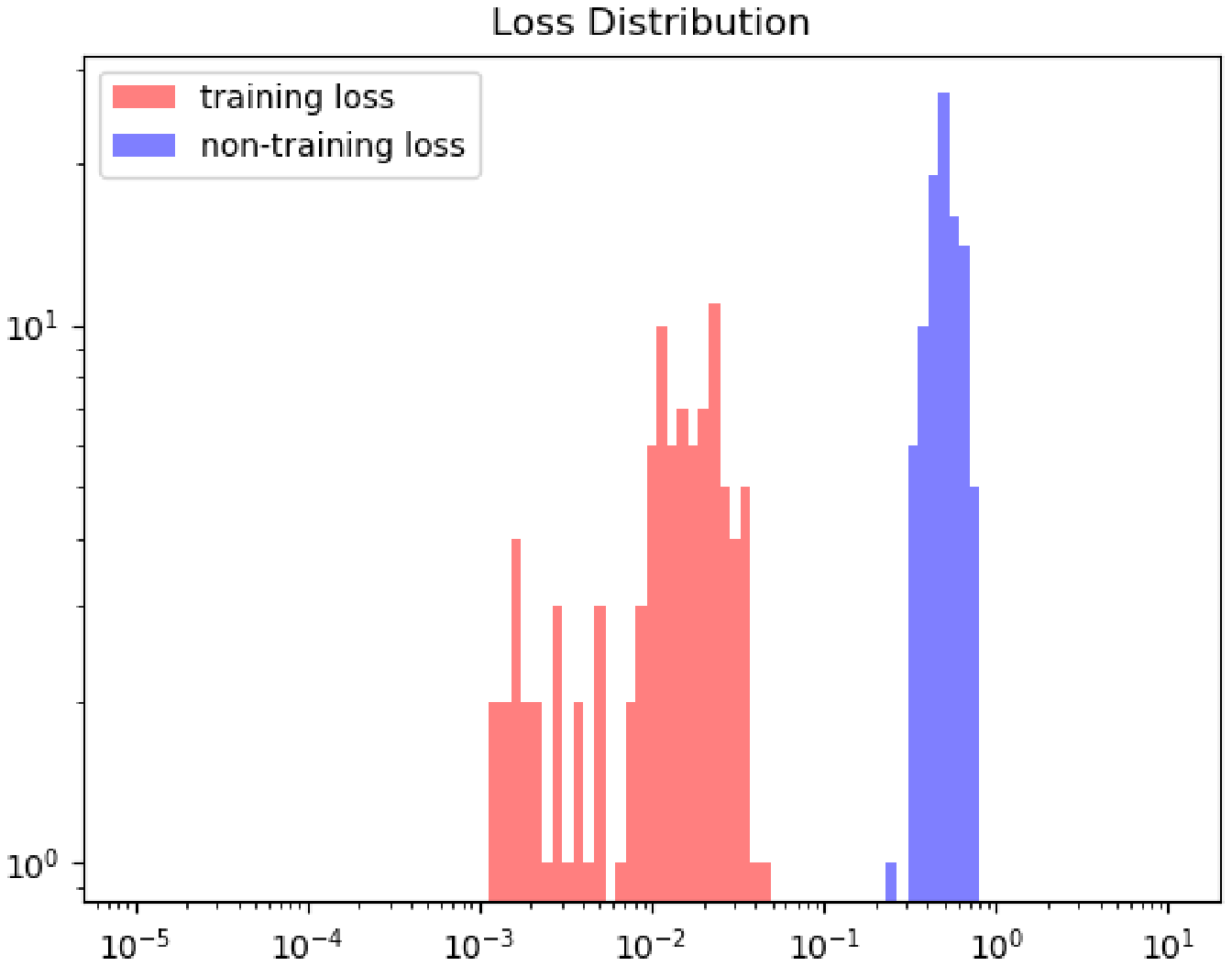}}
\subfigure[BadNets for $r=$ 8]{\includegraphics[width=.15\textwidth]{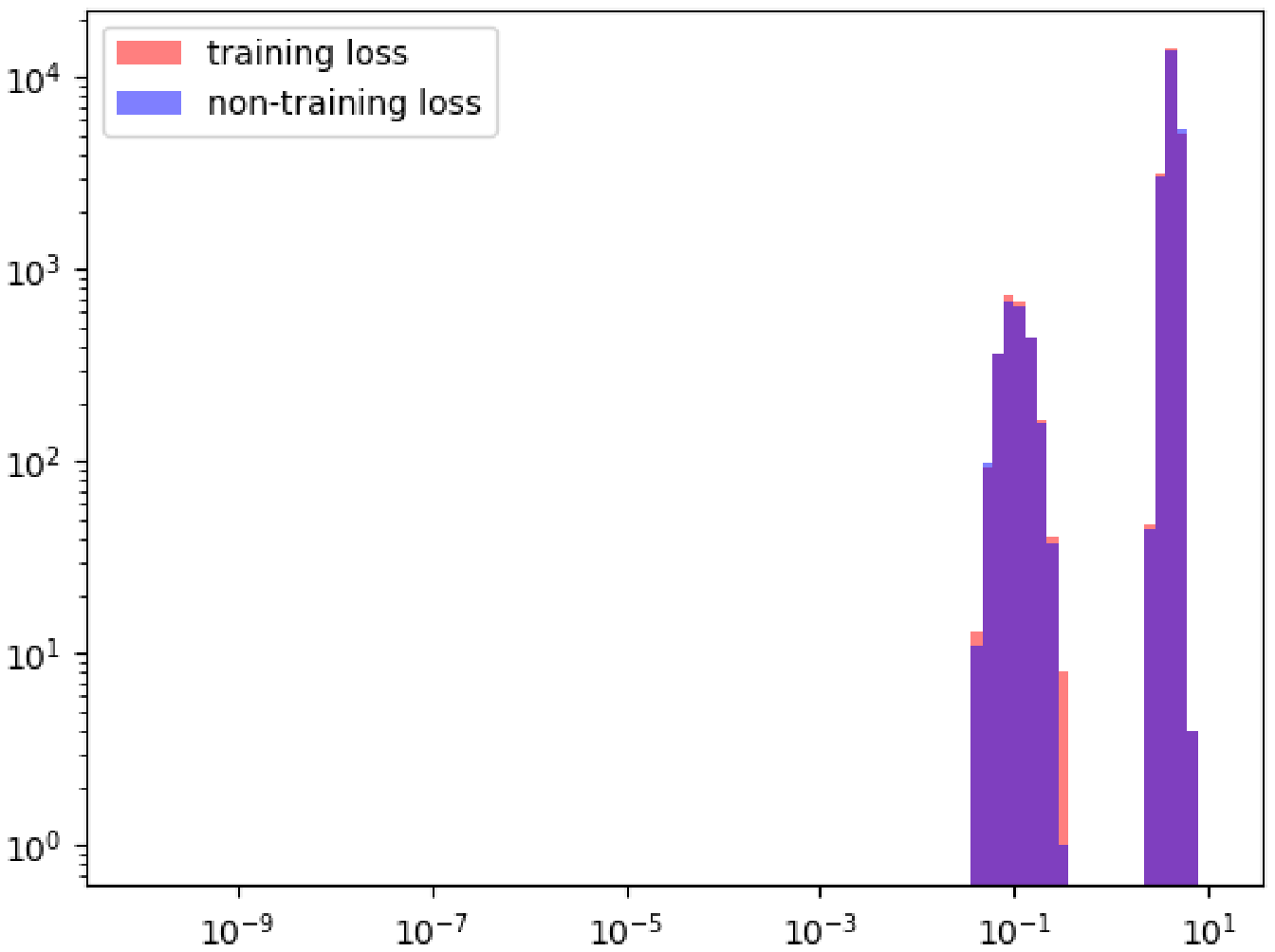}}
\subfigure[TaCT for $r=$ 8]{\includegraphics[width=.15\textwidth]{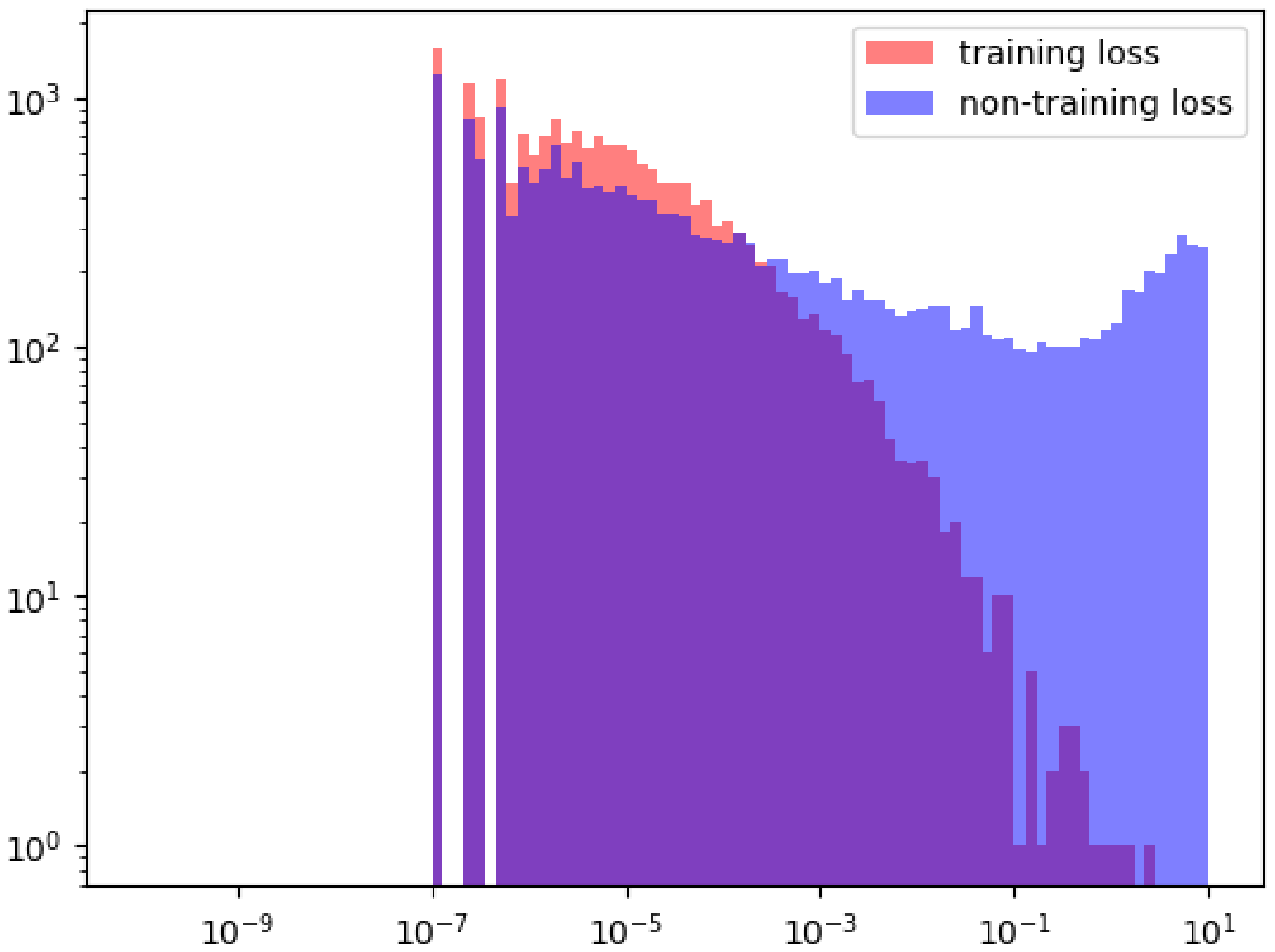}}
\subfigure[LIRA for $r=$ 8]{\includegraphics[width=.15\textwidth]{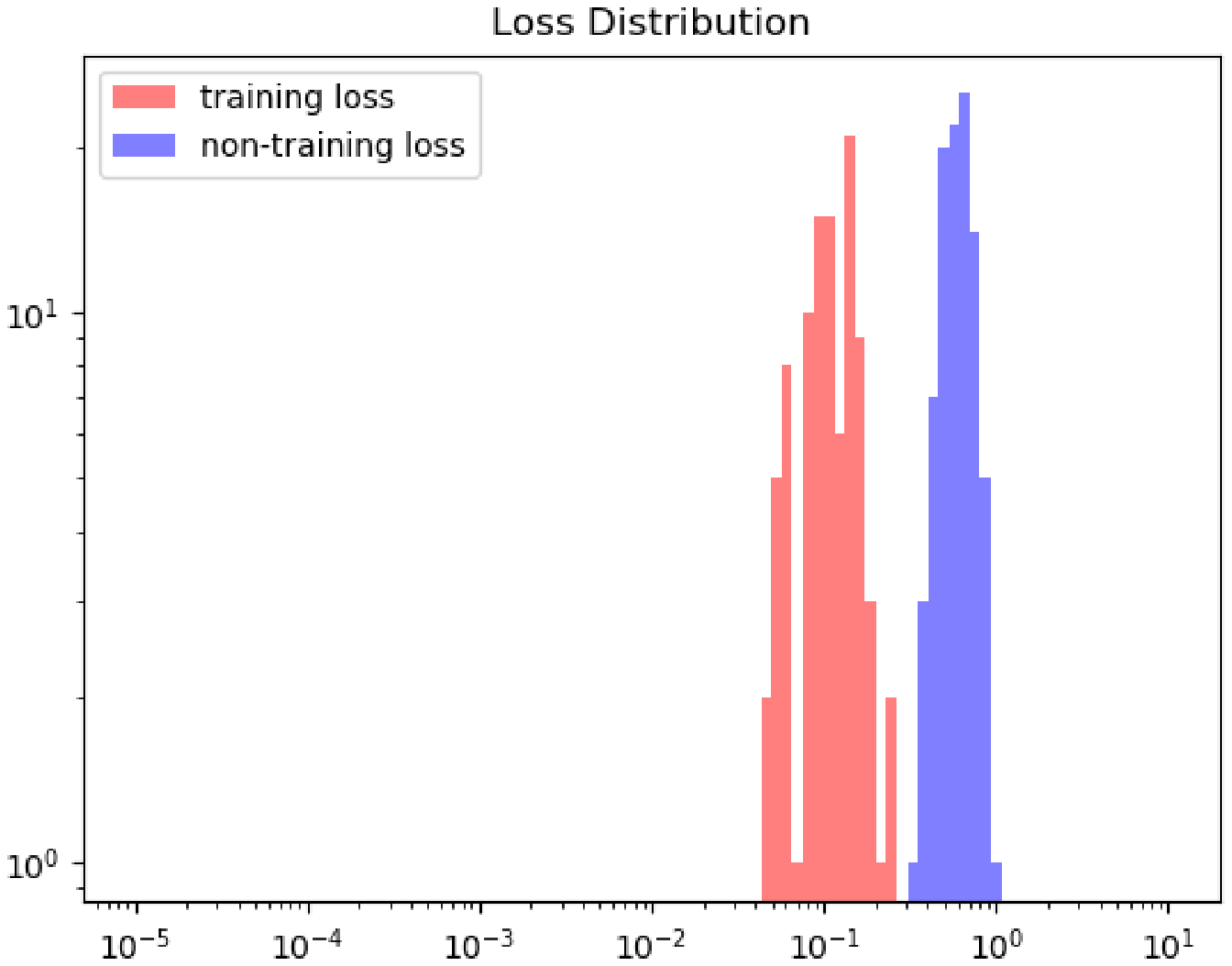}}
\subfigure[IBD for $r=$ 8]{\includegraphics[width=.15\textwidth]{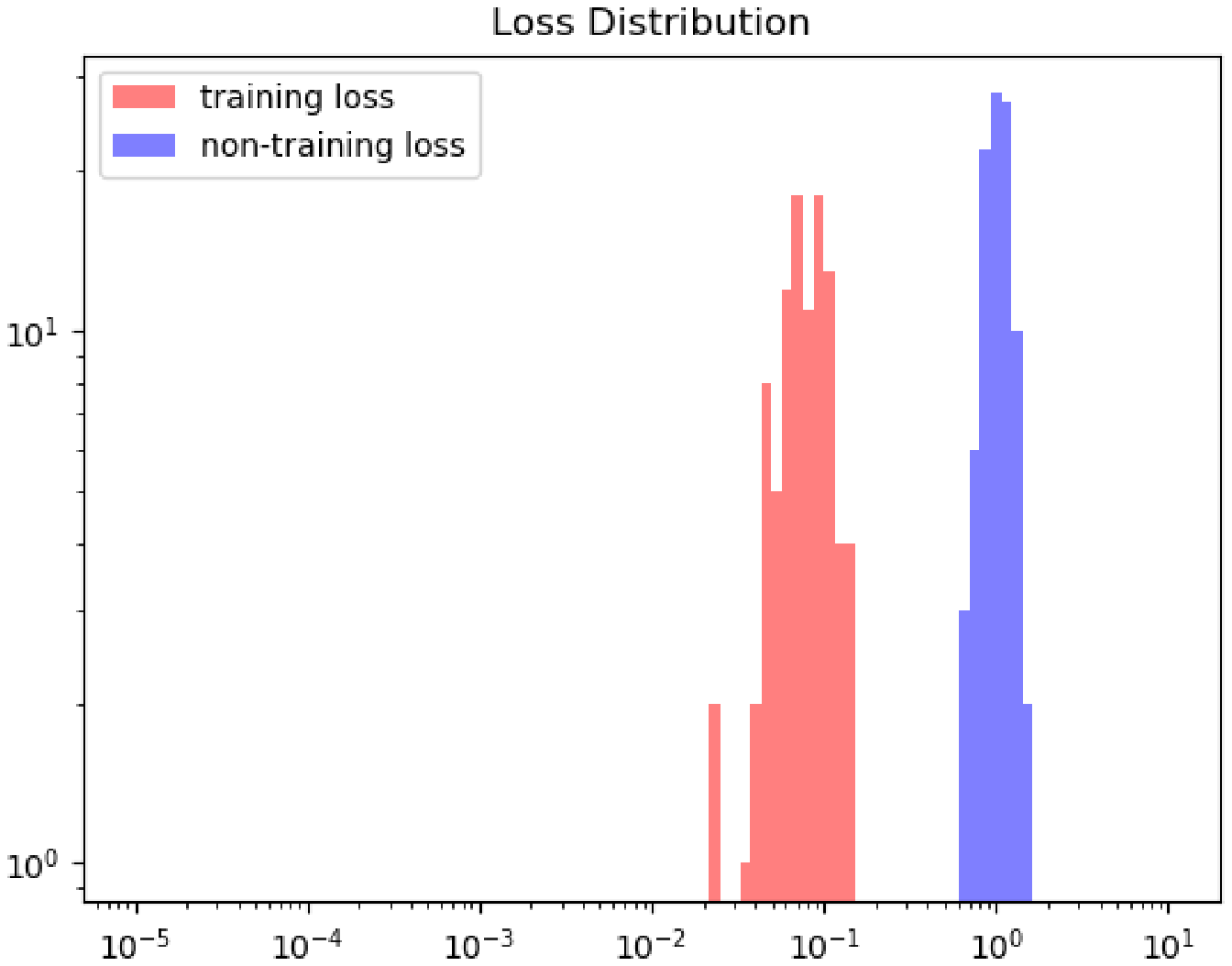}}
%%%%
\subfigure[Truth Serum for $r=$ 16]{\includegraphics[width=.15\textwidth]{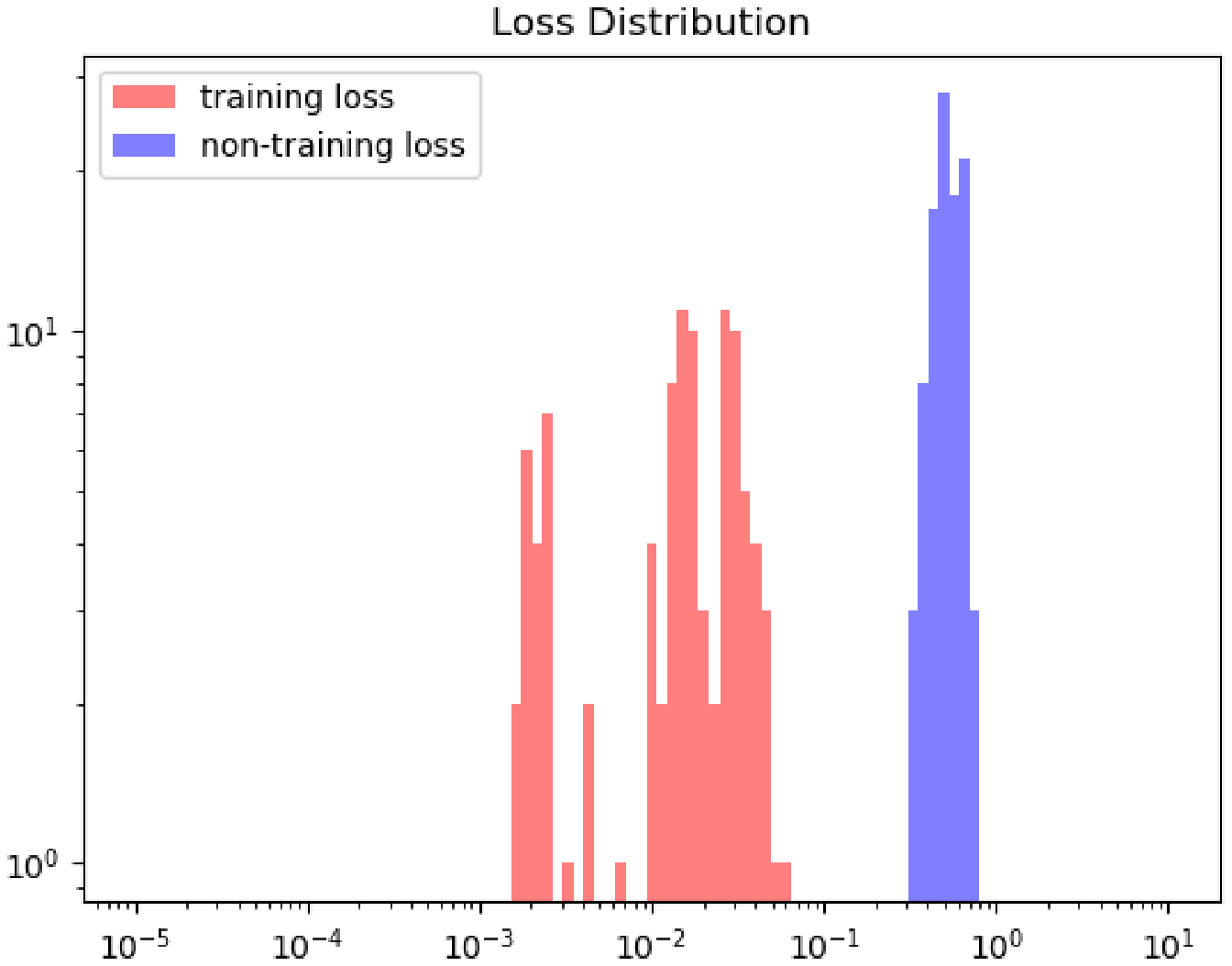}}
\subfigure[BadNets for $r=$ 16]{\includegraphics[width=.15\textwidth]{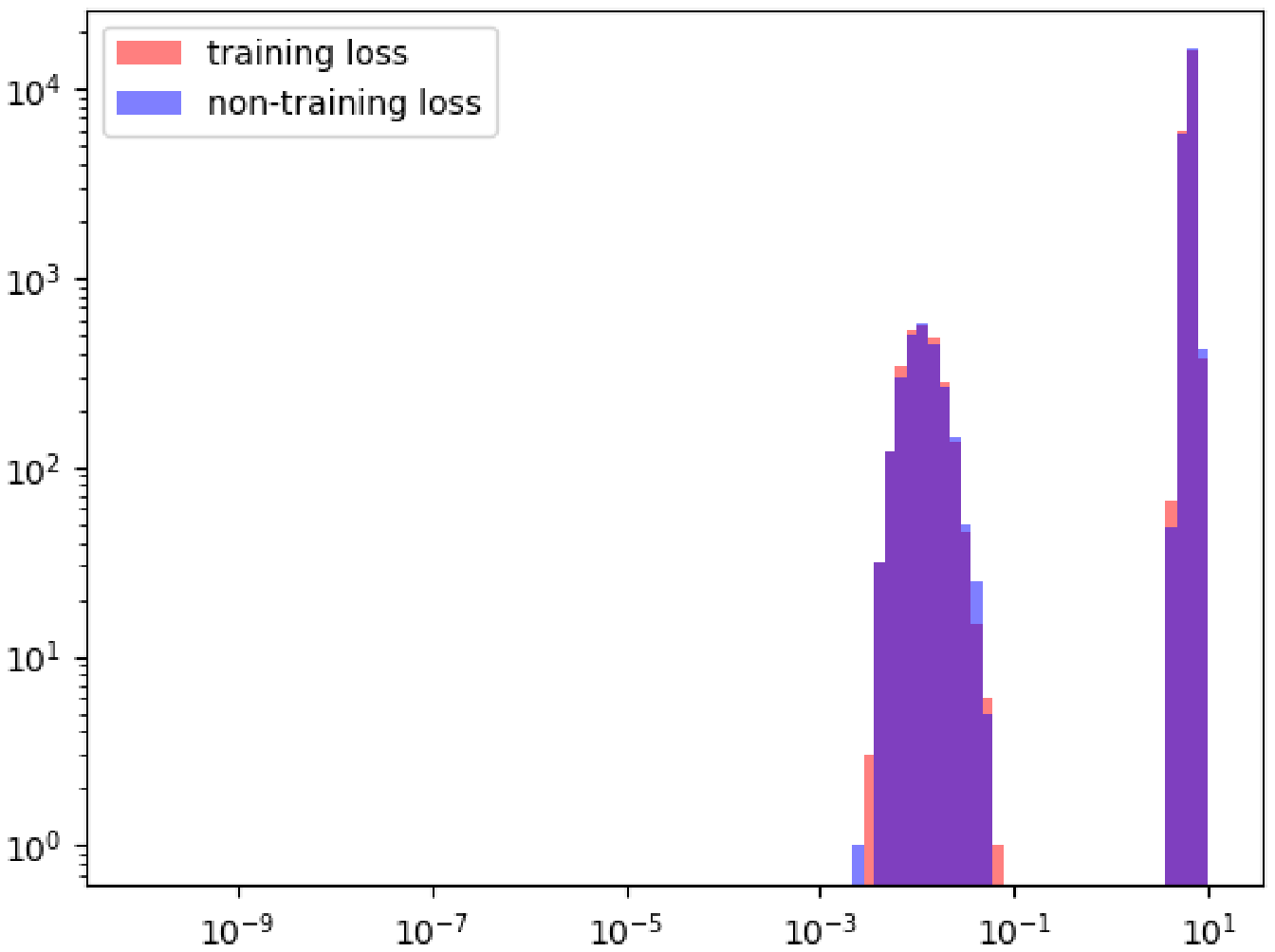}}
\subfigure[TaCT for $r=$ 16]{\includegraphics[width=.15\textwidth]{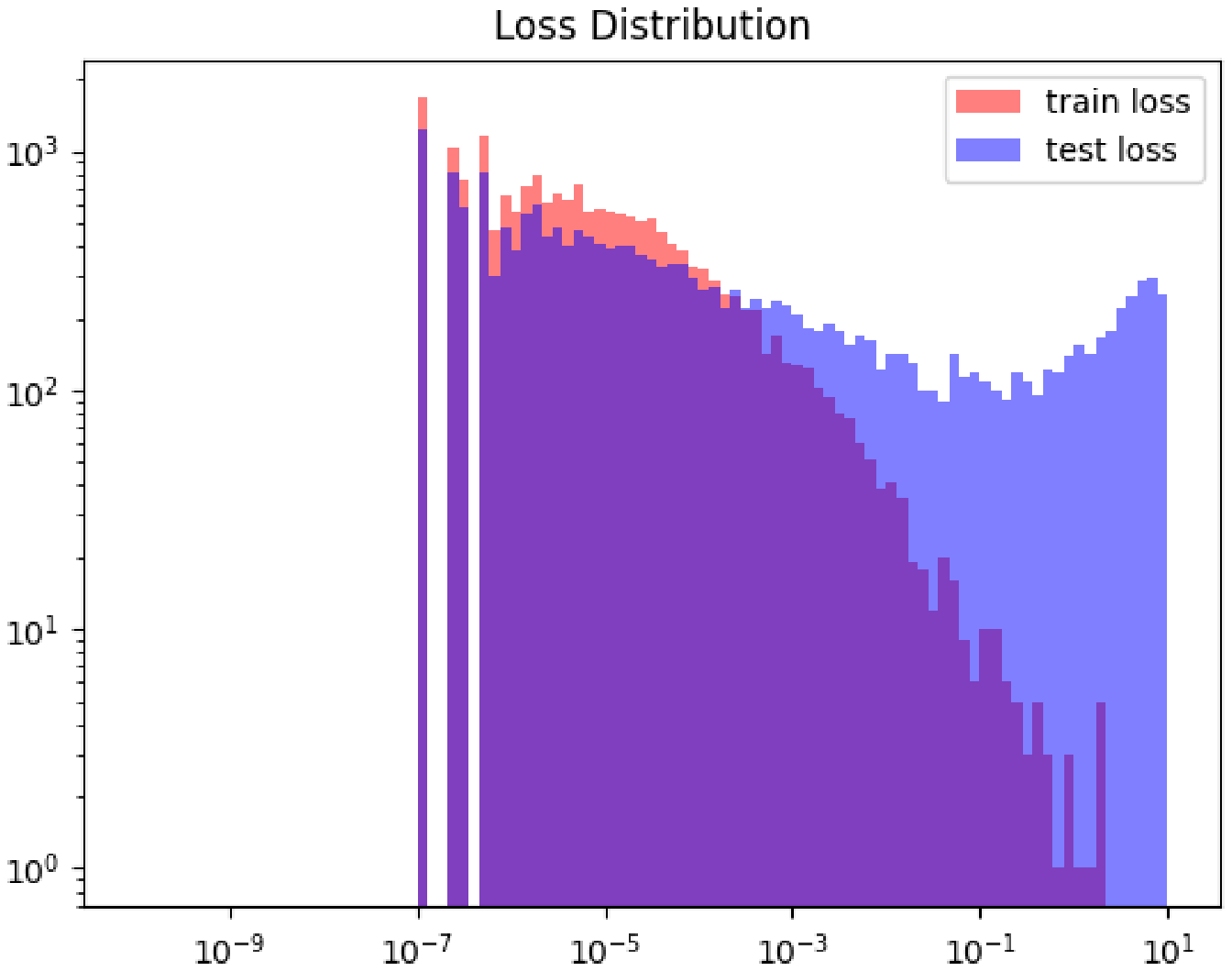}}
\subfigure[LIRA for $r=$ 16]{\includegraphics[width=.15\textwidth]{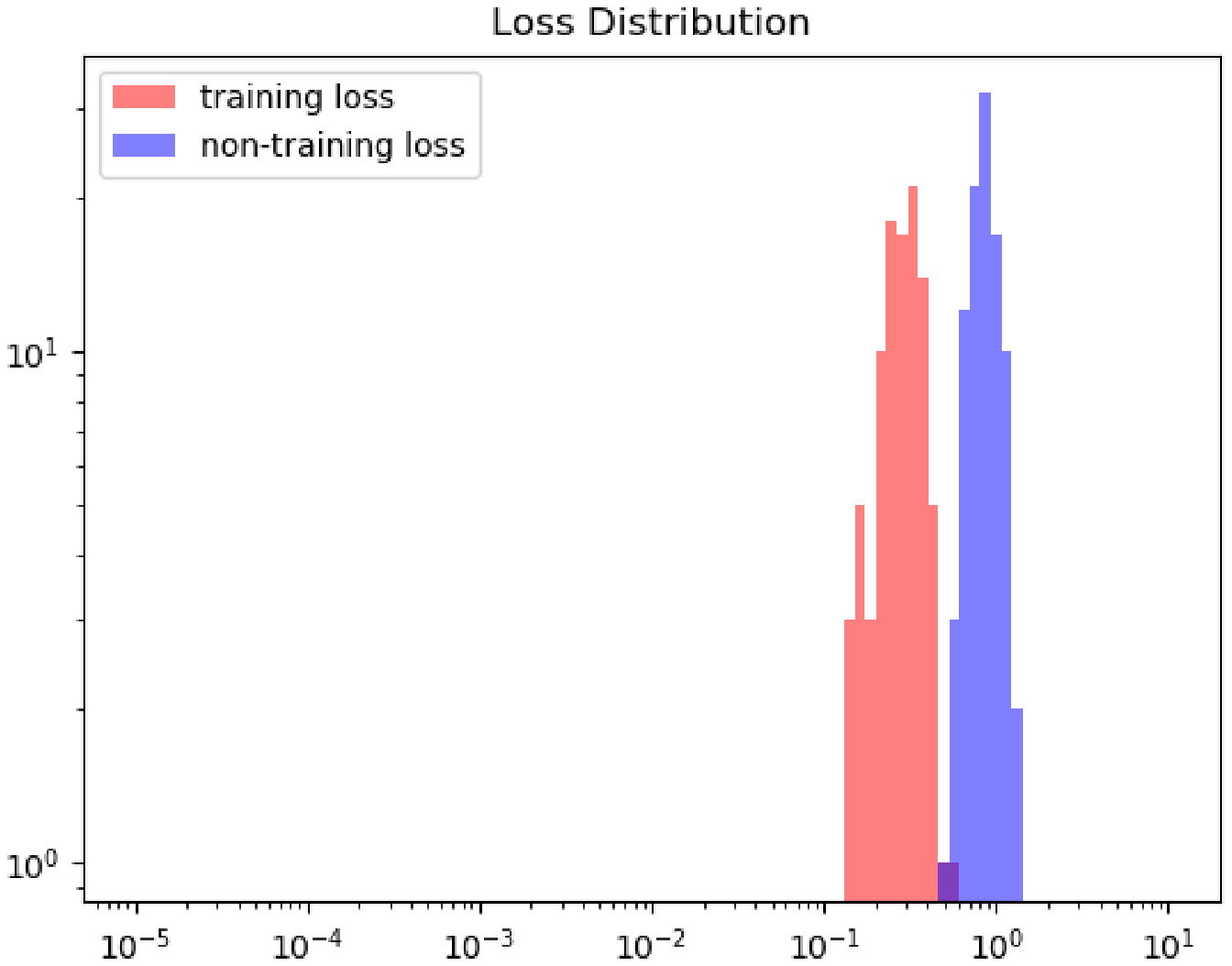}}
\subfigure[IBD for $r=$ 16]{\includegraphics[width=.15\textwidth]{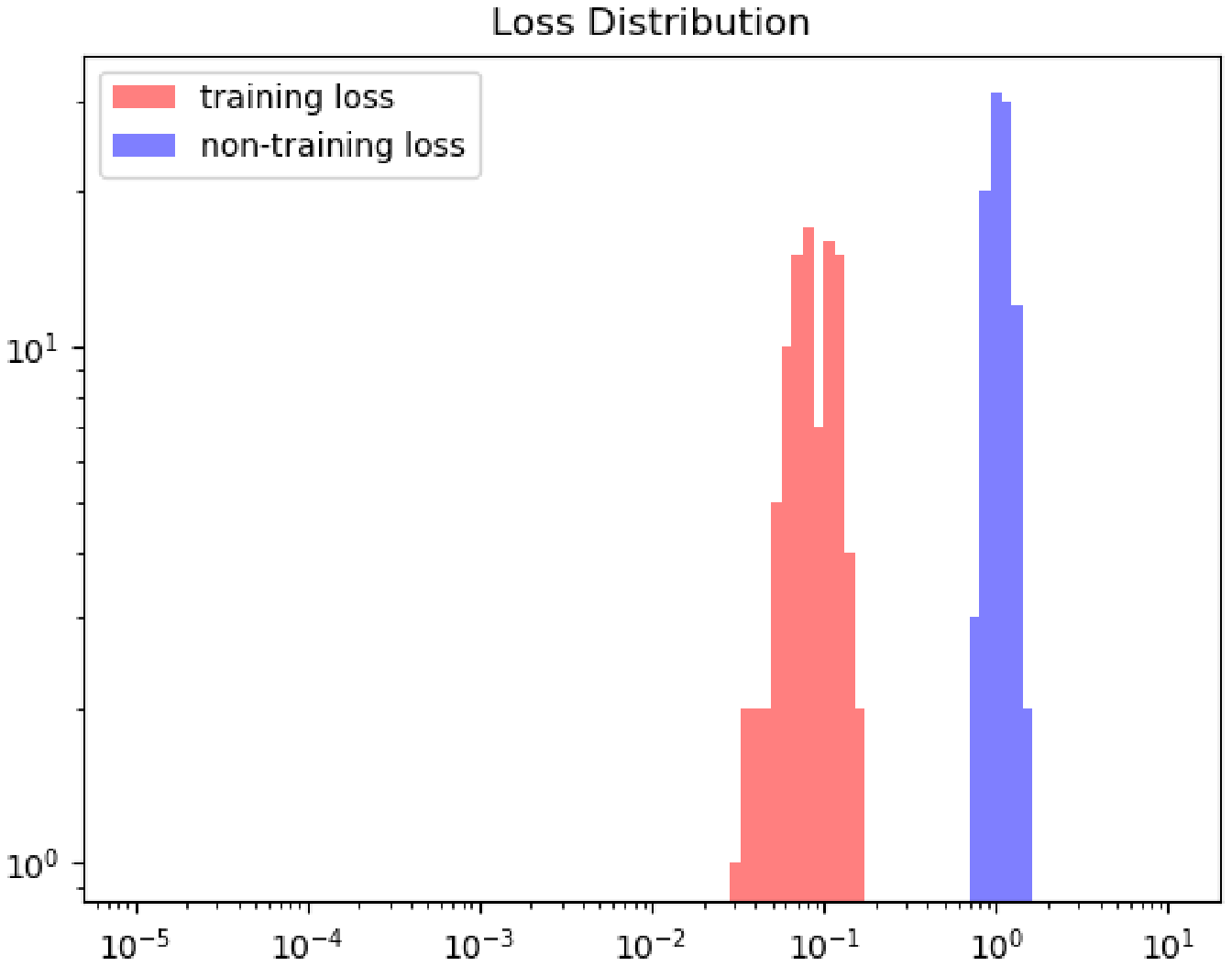}}
\end{subfigmatrix}
\caption{Loss Distributions on Victim Model for Each Attack}
\label{fig:loss_distribution}
Loss distributions in which each victim model infers its training data are red, and those in which the models infer its non-training data are blue. The number on each graph title varies with the replicated times of poisoning data.
\end{figure*}

Fig.~\ref{fig:loss_distribution} shows loss distributions of training and non-training samples for each model. 
According to the figure, Truth Serum boosts membership inference attacks because the losses of training and non-training samples are separated. 
Remarkably, borderlines between the losses of training and non-training samples are stable regardless of the size of $D_p$, and these phenomena are consistent with the original work~\cite{tramer2022truth} of Truth Serum. 

In contrast, BadNets and Tact are unable to separate the losses of training and non-training samples compared to Truth Serum. 
It is thus considered that MIA-SR and MIA-AUC deteriorate because the membership inference is almost random. 
More specifically, the loss distributions of training and non-training samples partially overlap. 
Then, MIA-SR and MIA-AUC should be improved compared to the Clean-Only model. 
Nevertheless, MIA-SR and MIA-AUC for TaCT are less than those of the Clean-Only model because the loss distributions of training and non-training samples are distributed over each other for most of a clean dataset $D$. 

Next, we discuss LIRA and IBD. 
According to Fig.~\ref{fig:loss_distribution}, their loss distributions of training and non-training samples are obliviously separated similarly to Truth Serum. 
We discuss why MIA-SR and MIA-AUC of LIRA and IBD deteriorate in the next section.

\subsection{Impact of Backdoors on Neuron Activations} \label{sec:non_assisted}

\begin{figure*}[h!]
\begin{subfigmatrix}{3}
\subfigure[Clean-Only]{\includegraphics[width=.30\textwidth]{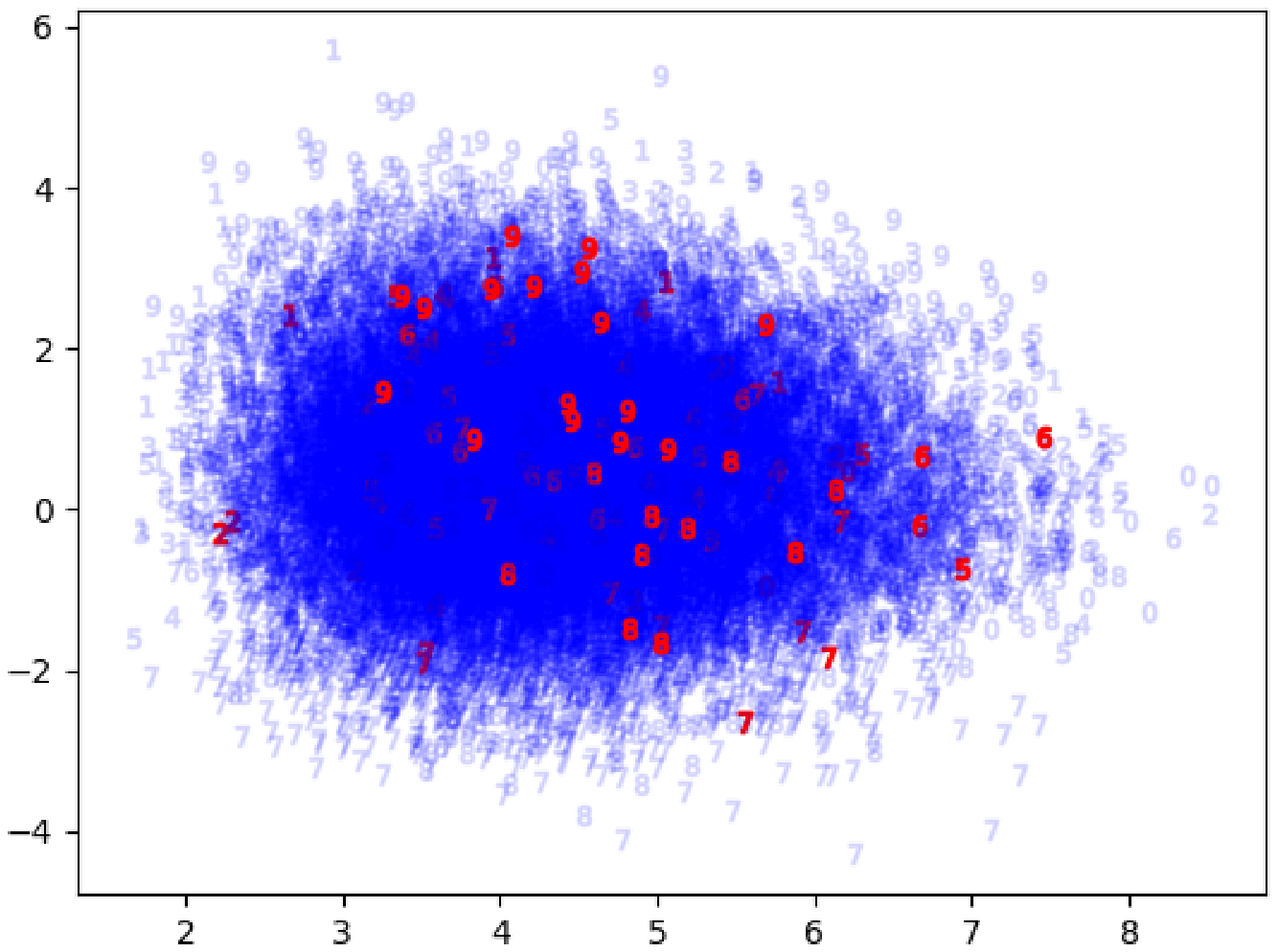}}
\subfigure[Truth Serum]{\includegraphics[width=.30\textwidth]{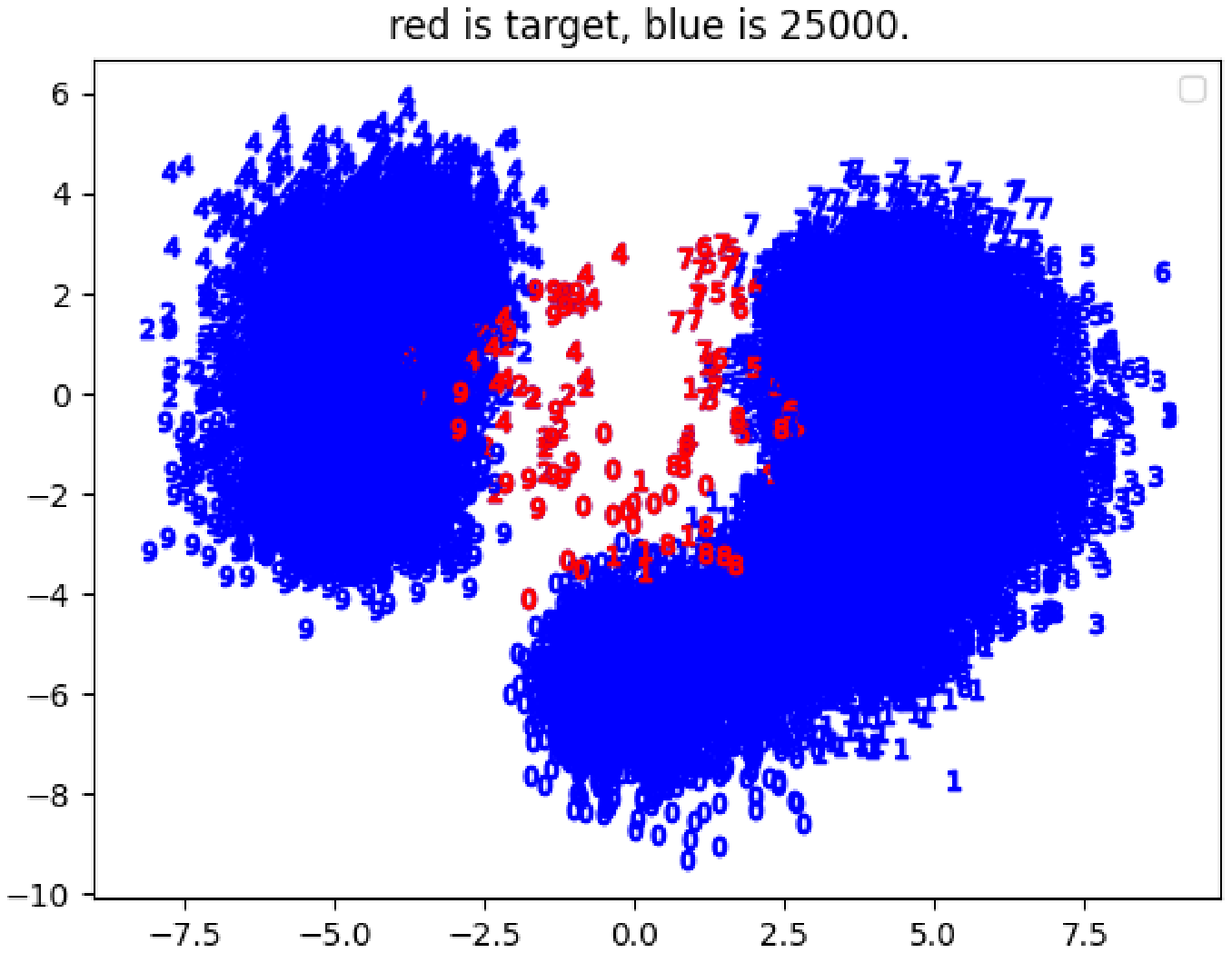}}
\subfigure[BadNets]{\includegraphics[width=.30\textwidth]{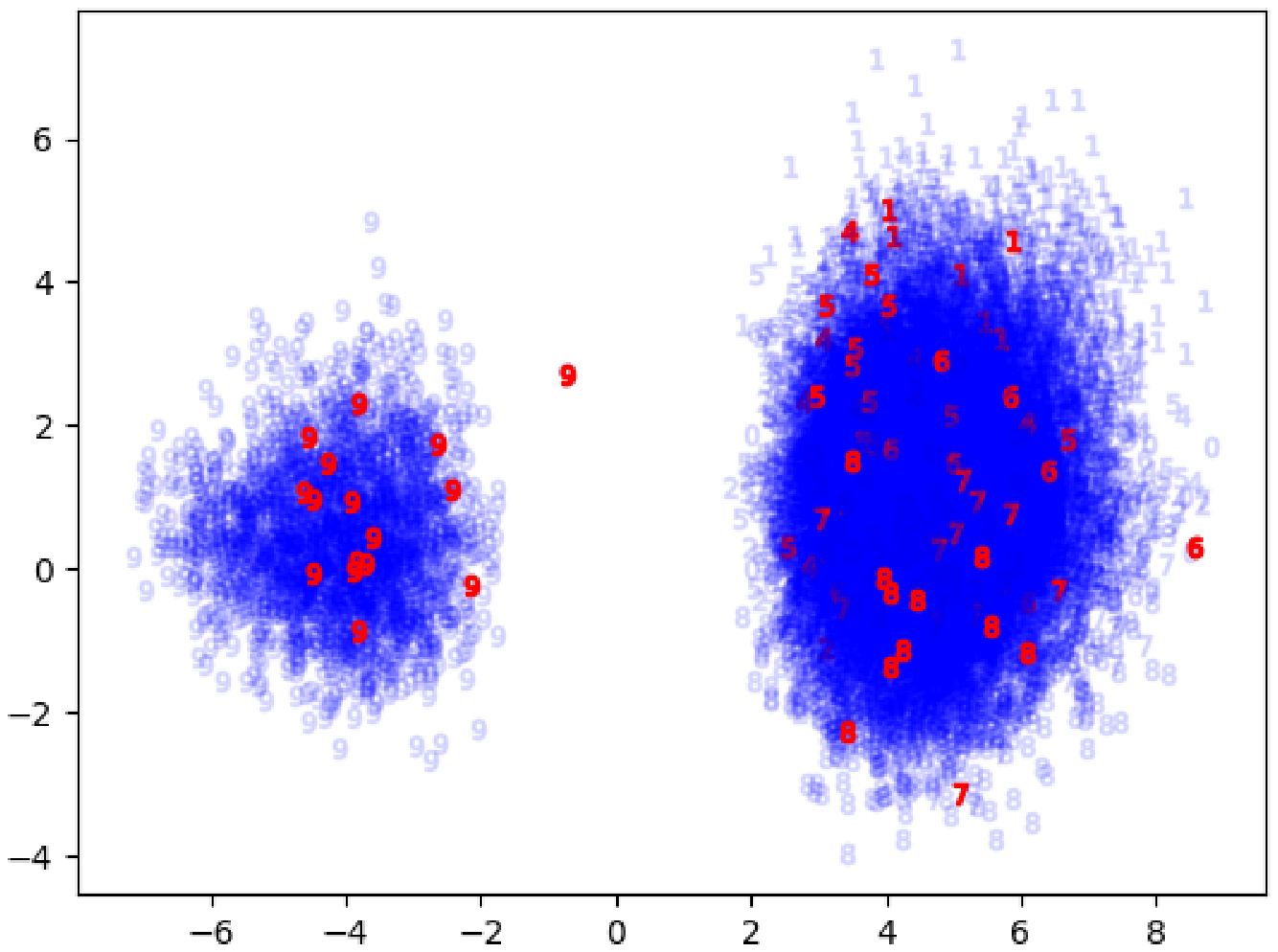}}
\subfigure[TaCT]{\includegraphics[width=.30\textwidth]{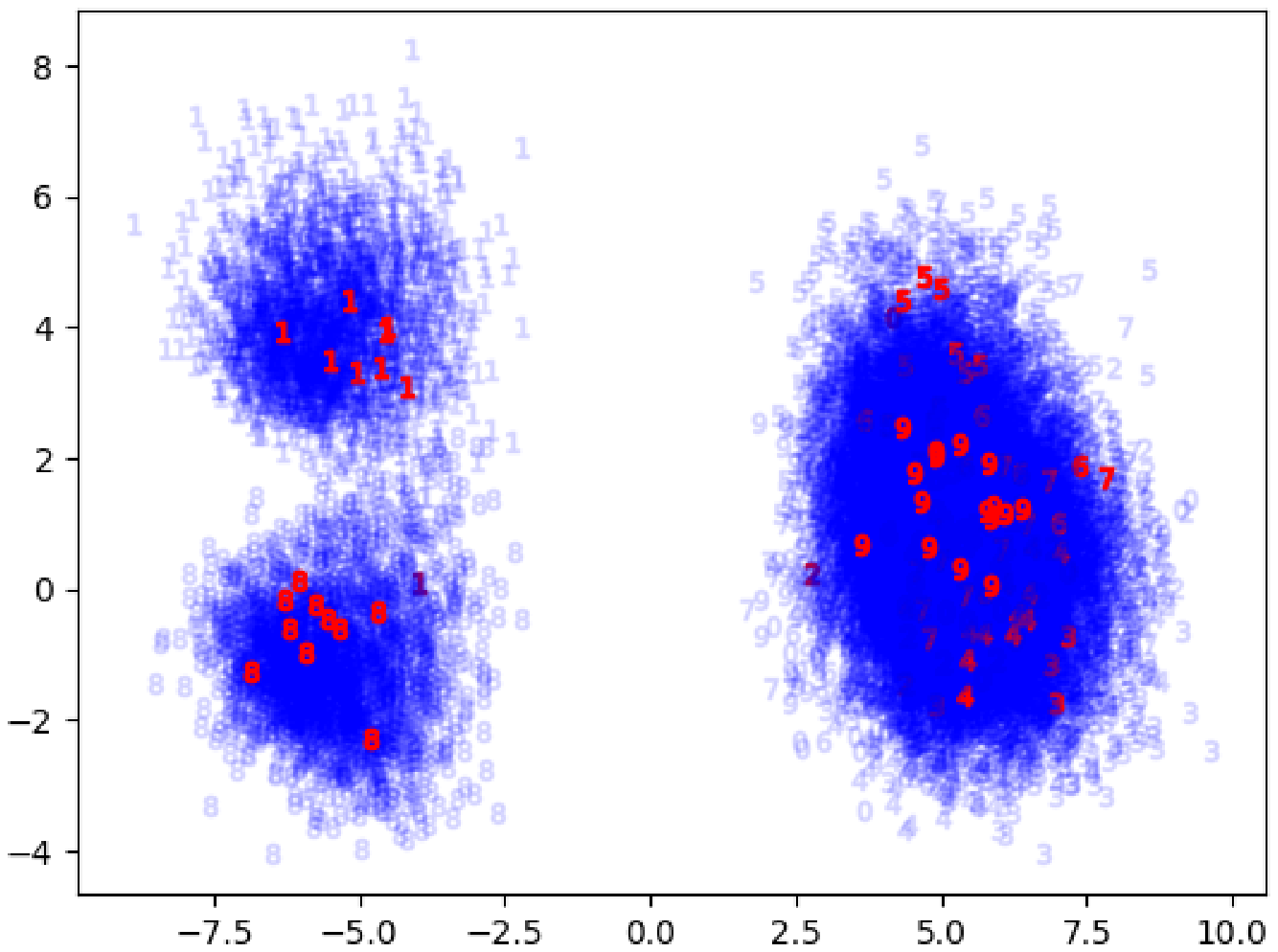}}
\subfigure[LIRA]{\includegraphics[width=.30\textwidth]{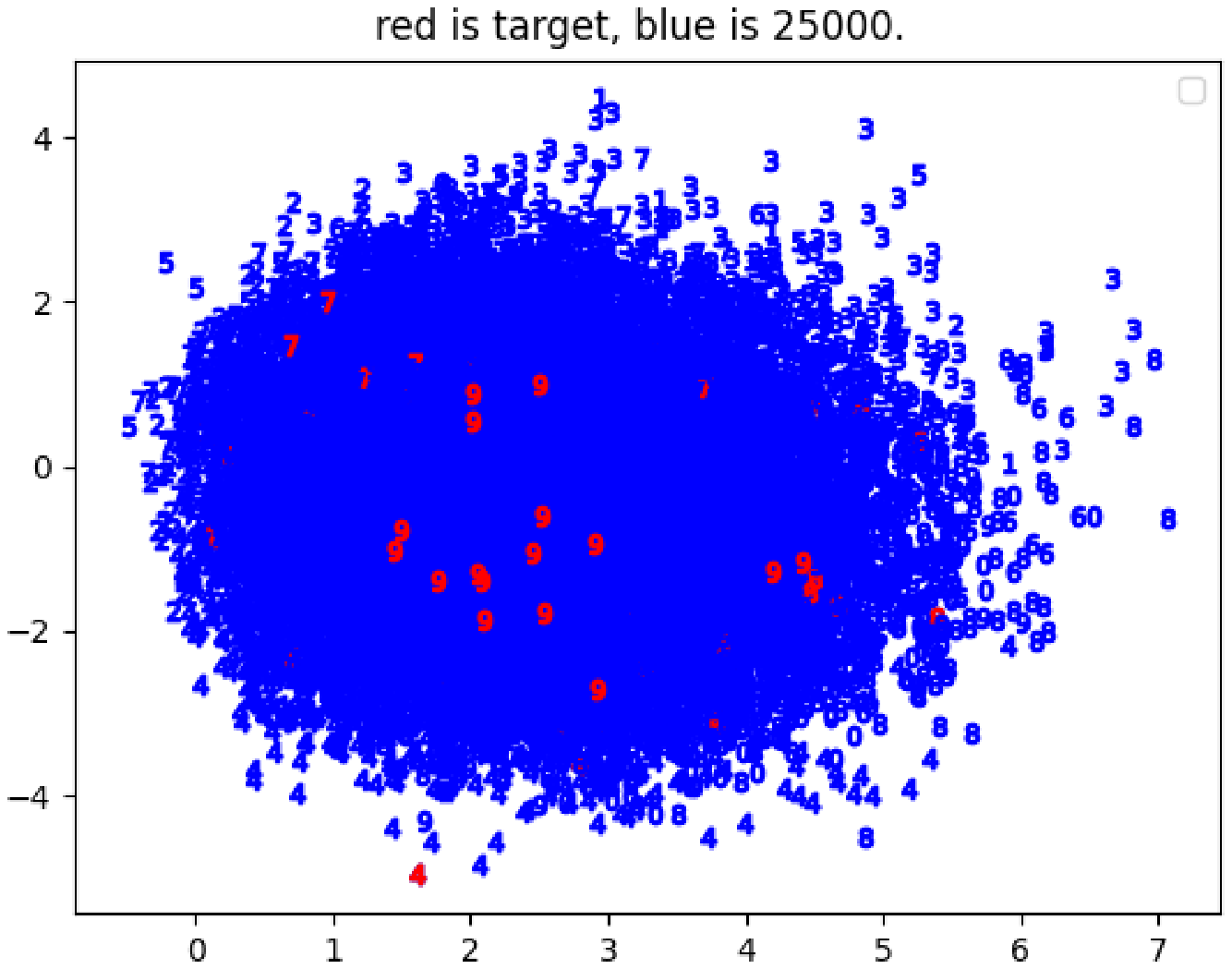}}
\subfigure[IBD]{\includegraphics[width=.30\textwidth]{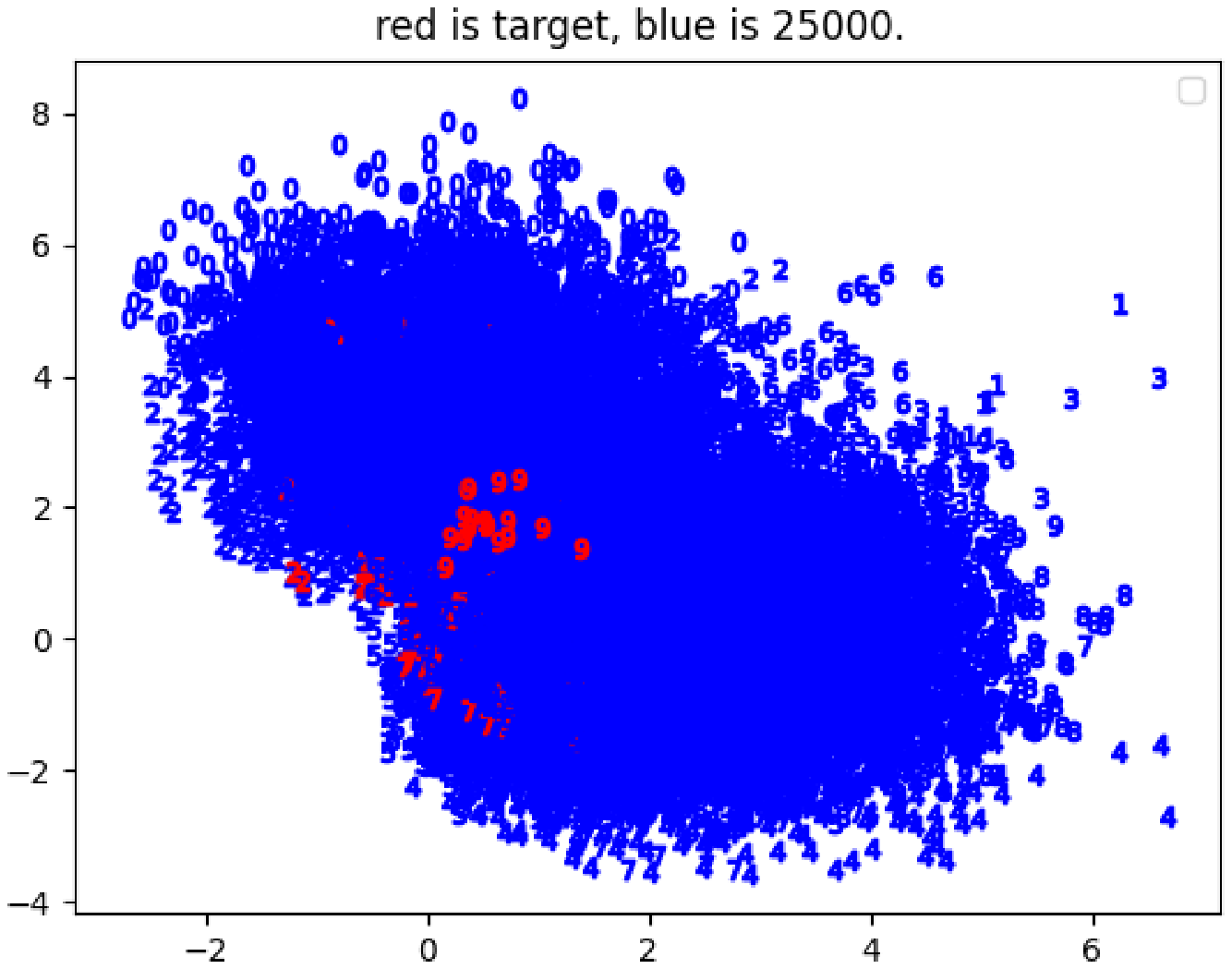}}
\end{subfigmatrix}
\caption{Neuron Activations on Training Data and Non-Training Data of Victim Model for Each Attack }
\label{fig:neuron_activation}
When each victim model infers training data or non-training data, neuron of the latent representation layer activates in this way. 4000 target data for membership inference attacks are plotted as red points, and 25000 training or non-training data are plotted as blue points.
\end{figure*}

We discuss the impact of backdoors on neuron activation from a standpoint different from the previous section to understand the backdoor-assisted membership inference attacks in more detail. 
In particular, we built a hypothesis that backdoors do not make loss distributions of training samples an outlier, and therefore a membership inference attack is unsuccessful even through the backdoor attacks. 
We observe neuron activations for each model to confirm the above hypothesis and then find strong evidence whereby a membership inference attack is unsuccessful. 
We show them in Fig.~\ref{fig:neuron_activation}. 

According to the figure, Truth Serum makes a target sample an outlier over the distribution of training samples, which is consistent with the original work~\cite{tramer2022truth}. 
%A poisoning attack causes it. 

By contrast, all the backdoor attacks make target samples inliers over the distributions of training samples. 
It means that triggers for backdoors have independent distributions of training samples. 
Namely, LIRA and IBD make target samples inliers, although LIRA and IBD separate loss distributions between training and non-training samples. 
It is thus considered that MIA-SR and MIA-AUC of LIRA and IBD deteriorate. 
The above phenomenon might be different if the test accuracies are improved. 

% loss distribution

% Activation Clustering (target 250 vs other)

\section{Conclusion}

We discussed backdoor-assisted membership inference attacks, which do not deteriorate the accuracy.
We first evaluated whether backdoor-assisted membership inference attacks with the original backdoors~\cite{Gu2019badnets} and the imperceptible backdoors~\cite{Tang2021demon,doan2021lira,zhong2022imperceptible} are successful in comparison with the existing poisoning-assisted membership inference attack~\cite{tramer2022truth}. 
%MIA-SR and MIA-AUC with various backdoors compared to the poisoning~\cite{tramer2022truth}.
We then showed that backdoor-assisted membership inference attacks are unsuccessful in contrast to the existing poisoning-assisted membership inference attack by Tramer et al.~\cite{tramer2022truth}

We also analyzed their resultant models with respect to loss distributions and neuron activations to deeply understand the reason for the unsuccessful results.
Then, we confirmed that triggers cannot affect the distribution of clean samples; namely, any clean sample becomes inliers while the existing attack makes it an outlier. 
%As a result, we confirm
Thus, we believe that backdoors cannot assist membership inference attacks.

\bibliographystyle{IEEEtran}
\bibliography{main}

\end{document}